\documentclass[manuscript]{aastex}
\usepackage{multirow}
\usepackage{lscape}
\begin{document}

\title{Constraining Galactic Magnetic Field Models with Starlight Polarimetry}
\author{Michael D. Pavel}
\affil{Institute for Astrophysical Research}
\affil{Boston University, 725 Commonwealth Ave, Boston, MA 02215}
\email{pavelmi@bu.edu}

\slugcomment{Accepted for Publication in The Astrophysical Journal - 20110712}

\shorttitle{Predicting Polarimetry}
\shortauthors{Pavel, M. D.}

\begin{abstract}

This paper provides testable predictions about starlight polarizations to constrain the geometry of the Galactic magnetic field, in particular the nature of the poloidal component. Galactic dynamo simulations and Galactic dust distributions from the literature are combined with a Stokes radiative transfer model to predict the observed polarizations and position angles of near-infrared starlight, assuming the light is polarized by aligned anisotropic dust grains. S0 and A0 magnetic field models and the role of magnetic pitch angle are all examined. All-sky predictions are made, and particular directions are identified as providing diagnostic power for discriminating among the models. Cumulative distribution functions of the normalized degree of polarization and plots of polarization position angle vs. Galactic latitude are proposed as tools for testing models against observations.

\end{abstract}


\section{Introduction}
A gap exists between theories of the large-scale Galactic magnetic field and observations that probe the same field. To bridge this gap, predictions from existing Galactic dynamo models that can be tested with current observational techniques, in particular polarization of background starlight, must be generated.

Extensive theoretical work over the past fifty years established the origin and expected nature of galaxy-scale dynamos. \citet{KZ08} provide a comprehensive review of cosmic magnetic field research and highlight many of the existing questions related to galactic dynamos. A leading theory for magnetic field generation in the Galaxy is the mean-field ``$\alpha$-$\Omega$'' disk dynamo \citep{M78,Ruz88}. This fundamental idea has led to many numerical simulations of Galactic magnetic fields \citep{E92,B92,B93,FS2000,K02,K03,HWK09,M10}. As these ideas have developed, they became more sophisticated and added much to the understanding of the Galactic magnetic field. Recent advances include understanding the role of magnetic helicity \citep{BS05,Sh06}, the interplay between the disk and halo fields \citep{MS08,M10}, and the first global MHD simulation of a cosmic ray-driven dynamo \citep{HWK09}.

The presence of magnetic fields in the Galaxy has been directly observed via several observational methods: Zeeman splitting of spectral lines \citep{V68,C93}, Faraday rotation of pulsars \citep{S68,Man74} and extragalactic sources \citep{CP62}, and synchrotron emission \citep{Wester62,Wie62,YZ84,YZ89}. Magnetic fields in the Galaxy have also been indirectly inferred from polarization of background starlight \citep{HI49,HA49,MF70} and polarized thermal dust emission \citep{H88}. Zeeman splitting and polarized dust emission only probe high-density regions in interstellar clouds and therefore only weakly contribute to our understanding of the large-scale structure and origin of the Galactic magnetic field. Polarized synchrotron emission has been used to study resolved magnetic fields in other galaxies \citep[and references therein]{B96} and in the Milky Way, in particular the vertical magnetic fields in the Galactic center \citep{YZ84}.

Much of the work attempting to understand large-scale magnetic fields in our Galaxy has used Faraday rotation of polarized emission from pulsars and extragalactic sources. Attempts have been made to fit empirical magnetic field models for the Galactic disk to the observed rotation measures \citep{RK89,MFH08,Sun08,NK10,VE11}. These attempts to measure basic properties of the Galactic magnetic field using rotation measures have proved to be challenging, particularly in light of arguments for an antisymmetric (odd) \citep{AM88,H97}, or symmetric (even) \citep{F01,Sun08} axisymmetric field in the disk, and for fields that cannot be so simply classified \citep{Mao2010}. Additional evidence suggests and that the observed rotation measure distribution may be strongly affected by nearby, small-scale structures \citep{W2010}.

Polarization of background starlight has previously been used to measure the large-scale orientation of the magnetic field \citep{H76}, to estimate the curvature and pitch angle (defined as arctan[$B_r/B_\phi$]) of the field \citep{H96}, and to study magnetic fields in other galaxies \citep{Hough96}. However, it has mostly been applied to the study of magnetic fields in and around star forming regions and individual clouds \citep[e.g.,][]{S85,G95,A98,T06,T07,K07}. By comparison, near-infrared (NIR) polarimetry of background starlight is a relatively untapped resource for the study of the large-scale Galactic magnetic field. Optical polarimetry has typically been restricted to probing magnetic fields within $\sim$1 kpc \citep{F02}, limiting its ability to probe the large-scale structure of the magnetic field. However, NIR polarimetry is less extincted by dust and can probe the magnetic field along longer lines of sight. The regular component of the Galactic magnetic field may be better measured by these longer sight lines \citep[longer than the scale length of the random component;][]{RK89,OS93} as the random magnetic components will tend to null averages. By studying the magnetic field in the outer Galaxy, where the dust density is lower than in the inner Galaxy \citep{SMB96,DS01}, and where there are fewer supernovae to create magnetic field disturbances \citep{H87}, NIR polarimetry may be used to probe the quiescent, regular Galactic magnetic field to multi-kpc distances.

Starlight polarimetry is insensitive to the directional polarity of magnetic fields, being sensitive only to their orientations. Therefore, starlight polarimetry cannot directly address questions about the nature of magnetic reversals seen in rotation measure studies. However, starlight polarimetry may shed light on the nature of the poloidal component of the Galactic magnetic field, complementing previous rotation measure studies \citep{TSS09,Mao2010}. By observing the field projected onto the plane of the sky, this method is sensitive to the Galaxy's poloidal magnetic field component and allows the nature of that field to be measured.

To advance the understanding of the Galactic magnetic field, strong constraints on theory are needed. In this paper, physics-based models of the Galactic dynamo, combined with models of the Galactic dust distribution, are used to make predictions for starlight polarizations that can be tested with current and emerging observational techniques. This will constrain the models, and therefore the physics, producing the Galactic magnetic field. The NIR stellar polarization mechanism and a Stokes radiative transfer model are outlined in \S 2. Section 3 presents the results of the NIR polarization simulations using the models described in \S 2. In \S 4, future comparison of these predictions with NIR polarimetry measurements and the constraints that can be placed on Galactic dynamo models are discussed. This work is summarized in \S 5.

\section{Simulating NIR Polarimetry}

This section describes the assumptions used in simulating NIR stellar polarizations. The polarization mechanism is briefly discussed, as the focus of this paper is more about the observables and less so about the physical mechanisms causing polarization. The specific magnetic field and dust models are listed, followed by a brief discussion of the properties of the simulated stellar populations. Finally, the Stokes radiative transfer model is presented.

\subsection{Polarization Mechanism}

Polarization of background starlight is thought to arise from anisotropic dust grains aligned in magnetic fields \citep[for a comprehensive history of this idea see][]{Laz03}. Theories for alignment mechanisms include paramagnetic dissipation \citep{DG51,P75,P79}, mechanical (gas streaming) alignment \citep{G52}, superparamagnetic dissipation \citep{JS67}, and radiative torques \citep{DM76,DW96,DW97,LH07}. With the exception of mechanical alignment, the prediction is that spinning dust grains will preferentially align with their long axes perpendicular to the local magnetic field. Unpolarized background starlight passing through a medium with aligned grains will experience a larger extinction cross-section perpendicular to the magnetic field than parallel to it, so that the transmitted light becomes weakly linearly polarized parallel to the direction of the magnetic field as projected onto the plane of the sky\footnote[1]{Throughout this work, polarization position angles will be measured in the Galactic coordinate system.}

\subsection{Magnetic \& Dust Models}

Two broad classes of Galactic magnetic fields need to be considered: axisymmetric (ASS) and bisymmetric (BSS), though higher-order symmetries and combinations are possible \citep{WK93}. ASS and BSS fields are differentiated by their symmetry with respect to the rotation axis \citep{KHB89} and both classes have been observed in other galaxies \citep{B96}. Galactic magnetic fields are further distinguished by their disk symmetry, either symmetric (even) or antisymmetric (odd) across the midplane \citetext{\citealp{KHB89}; see Table 1 in \citealp{B96}}. Here, only ASS magnetic field models are considered; these are classified as A0 (odd) or S0 (even). This work will also consider disk-even, halo-odd (DEHO) magnetic fields which have observational \citep{Sun08} and theoretical \citep{M10} foundations.

\begin{figure}
	\includegraphics{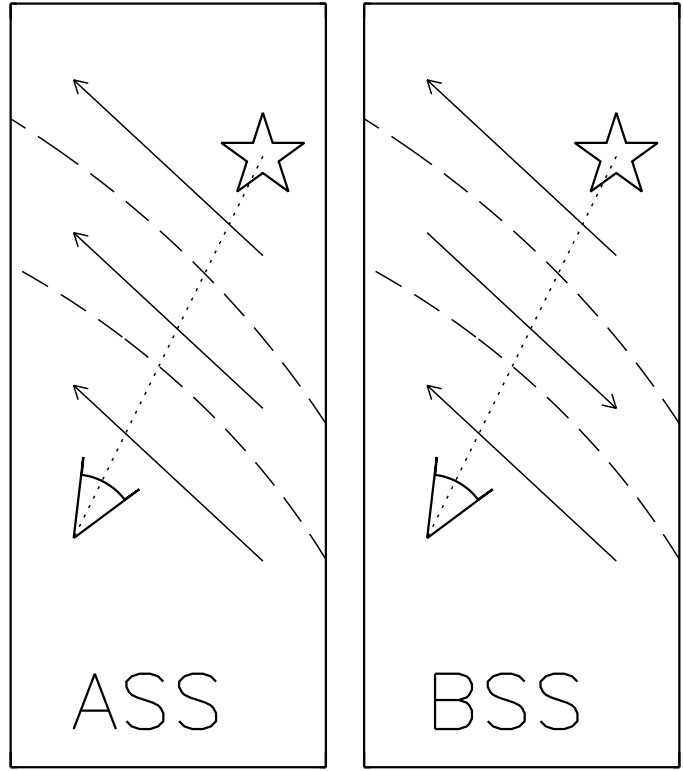}
	\centering
	\caption{\label{cartoon} Cartoon illustrating the insensitivity of polarization of background starlight to magnetic polarity reversals, seen in a top-down view of a galaxy. A star (star symbol) emits light that passes through a spiral arm (deliniated by dashed lines) to an observer (eye symbol) for the cases of an axisymmetric ASS (left) and bisymmetric BSS (right) magnetic field. The arrows indicate the directional polarity of the magnetic field in each region. The observer would measure identical linear starlight polarization properties for the star in both cases.}
\end{figure}

The polarizing mechanism described here is sensitive not to the directional polarity of the magnetic field, but to its orientation \citep{H96}. Consider the two simple magnetic field geometries in Fig. \ref{cartoon}: one in which the magnetic field inside a spiral arm has the same directional polarity as the magnetic field outside the arm (ASS magnetic field), and one in which the magnetic field inside the arm has the opposite directional polarity of the rest of the Galaxy (BSS magnetic field). A distant star whose light passes through this spiral arm to an observer would gain identical polarizations in both cases because the orientation of the fields is identical even though the directional polarities may differ. Ergo, background starlight polarimetry is insensitive to magnetic field reversals.

The predicted NIR polarizations were generated using simple magnetic fields configurations: axisymmeteric analytic magnetic fields; and axisymmetric dynamo-generated configurations from the literature, in particular the dynamo simulations of \citet{FS2000} and \citet{M10}. From \citet{FS2000}, the numerical outputs of three S0 and three A0 magnetic field models were obtained (D. Schmitt, private communication). These magnetic field models had magnetic pitch angles from $7.0-10.6\degr$ and are listed in Table \ref{model_table}. From \citet{M10}, the numerical outputs of three DEHO magnetic field models (513b, 511b, and 527b) were obtained (D. Moss, private communication), corresponding to the three models shown in Fig. 4 of that paper. This current work focuses on using these dynamo simulations as physically-motivated representatives of A0, S0, and DEHO Galactic magnetic fields. Three analytic, axisymmetric models (these can also be classified as S0 models) were also considered: one consisting of purely azimuthal magnetic fields ($B_r=B_z=0$) and two possessing azimuthal and radial fields (but no $B_z$ component) with magnetic pitch angles matching those found for recent models in the literature of $11.5\degr$ \citep{B07} and $24\degr$ \citep{R10}. Since polarization of background starlight is not sensitive to the directional polarity of magnetic fields the simulations used do not include magnetic reversals.

Two different dust models \citep{SMB96, DS01} of the Galactic dust distribution were used, both derived from NIR and FIR emission measured by the COBE/DIRBE instrument for Galactic latitudes $|b| < 30\degr$. Since the observations simulated here extend to $|b| \geq 30\degr$, the exponential forms utilized in the dust models were assumed to adequately describe the Galactic dust distribution far from the midplane. \citet{SMB96} imposed axisymmetry, with fixed disk scale length and scale height values. \citet{DS01} invoked a more sophisticated model which included an axisymmetric disk with a linear flair, the local arm, four spiral arms, and a Galactic warp.

\subsection{Stellar Population Sample}

Observed background starlight polarizations contain contributions from all aligned dust along the line of sight to the stars, so stars at different distances may sample different polarizing media and exhibit different polarization properties. Therefore, to accurately simulate real observations, the background stellar sources used here must be reasonably distributed along and across Galactic lines of sight. The stellar population model\footnote[2]{http://model.obs-besancon.fr/} of \citet{R03} has been shown to agree with NIR star counts from 2MASS \citep{Re09}, so this model was used to generate $10\arcmin \times 10\arcmin$ simulated star fields across the entire sky. Monte-Carlo selections of stars were drawn from extinction-free stellar probability distributions. Stellar H-band (1.6$\mu$m) photometry for each star was calculated using Johnson-Cousins filter transmission curves and were extincted assuming a standard diffuse Galactic extinction of 1.8 mag kpc$^{-1}$ at V-band \citep{W03} in the disk. The stellar distribution model returned stellar types, atmosphere parameters, mass, distances, Johnson-Cousins apparent magnitudes, apparent colors, and visual extinctions. Using these stellar samples, starlight polarization predictions were made for the entire sky, spaced every five degrees in Galactic latitude and longitude. To simulate realistic observational constraints, stars dimmer than $H=14$ from the Monte-Carlo stellar population samples were excluded. 

\subsection{Radiative Transfer Model}

The propagation of light from an unpolarized background star to an observer through aligned dust grains imposes weak linear polarization. To calculate the polarization observables P (the degree of polarization) and PA (the Galactic position angle of polarization, measured toward increasing Galactic longitude from the Galactic North pole), unpolarized light was modeled as leaving each simulated star and propagating to the observer. As the starlight propagates, the changes in the Stokes I, U, and Q parameters were calculated \citep{M74}. All simulations remained in the optically thin regime with no contribution from polarized emission from any dust, as is appropriate for NIR wavelength observations of the Galactic disk. This assumption may not apply within $1\degr$ of the Galactic plane where the dust optical depth can become large.

To illustrate aspects of the radiative transfer model, consider two limiting cases: a line of sight through two distinct polarizing regions whose magnetic fields are aligned with the same orientation, and a line of sight through two polarizing regions whose magnetic field orientation on the sky are aligned perpendicular to each other. In the first case, as the initially unpolarized background starlight passes through the first region a weak linear polarization (along the orientation of the magnetic field) is imposed. The light passes through the second polarizing region which further polarizes the background starlight along the same orientation. In the second case, the unpolarized background starlight again gains polarization from the first polarizing region. However, the second region \textit{depolarizes} the starlight and the observer sees much weaker or unpolarized starlight. Because of the possible depolarizing effect of multiple polarizing regions, a single star may be a poor probe of the full nature of the magnetic field along a line of sight. Therefore, this work uses ensembles of stars (distributed with distance) to characterize each simulated sky direction.

The Sun is assumed to be located 8.5 kpc from the Galactic center and in the Galactic midplane. For each simulated star, the entire line-of-sight from the star to the Sun was divided into 10 parsec blocks. This size was chosen to be smaller than the typical variations found in any of the magnetic and dust models. At each block, the local dust density was calculated, the magnetic field direction from the models (as seen from the Sun's position) was projected onto the sky, and the changes in Stokes parameters were calculated by integrating through that block. The changes in the Stokes parameters, from \citet{LD85}, are described by the equations:
\begin{equation}
	\frac{d(Q/I)}{ds} = \epsilon \: \rho \: \pi \: a^2 \: cos \: 2 \psi \: cos^2\gamma \: \frac{(Q_\bot - Q_{||})}{2} ,
\end{equation}
\begin{equation}
	\frac{d(U/I)}{ds} = \epsilon \: \rho \: \pi \: a^2 \: sin \: 2 \psi \: cos^2\gamma \: \frac{(Q_\bot - Q_{||})}{2} ,
\end{equation}
\begin{equation}
	\frac{dI}{ds} = -I \: \epsilon \: \rho \: \pi \: a^2 \: \frac{(Q_{||} + Q_\bot)}{2}
\end{equation}
where $\epsilon$ is the polarizing efficiency, $\rho$ is the local dust density, a is the effective radius of the dust grains (assumed to be 0.1$\mu$m `astronomical silicate'), $\psi$ is the direction of the magnetic field projected onto the plane of the sky, $\gamma$ is the angle between the direction of propagation of the radiation and the local magnetic field direction, and Q is the scattering efficiency of the long (perpendicular) and short (parallel) axes from \citet{D85}. The polarizing efficiency of the dust is assumed to be identical throughout the Galaxy.

The degree of polarization and position angle (PA) can then be calculated:
\begin{equation}
	P = \sqrt{U^2 + Q^2}
\end{equation}
\begin{equation}
	PA = \frac{1}{2}\arctan(U/Q)
\end{equation}
The uncertainty, and possible variation, of the polarizing efficiency of dust in the Galaxy affect the calculation of the absolute degree of polarization. Instead, the distributions of modeled polarizations were post facto normalized by the maximum calculated polarization in order to facilitate comparison with observed NIR polarization distributions. No additional constraints have been placed on the simulated star parameters, and all stars with $H<14$ are used in the following analysis.

Simulations for all 24 models were generated across the entire sky for five degree steps in Galactic latitude and longitude. Each of the 24 models consists of one magnetic field model plus one Galactic dust model. The model designations and their associated magnetic field and dust models are listed in Table \ref{model_table}. Model numbers 1-6 are the \citet{FS2000} S0 models; models 7-12 are the \citet{FS2000} A0 models; models 13-18 are the \citet{M10} DEHO models; models 19-24 are the axisymmetric analytic models. For each model, the P and PA for each star in each Monte-Carlo stellar sample was calculated. The mean PA and its dispersion were calculated for one $10\arcmin \times 10\arcmin$ field of view at each sky grid direction for each model. Cumulative distribution functions (CDFs) of the polarizations from each model were tabulated. The polarization CDFs are shown in Fig. \ref{CDFs} and the quartile values of the cumulative $P/P_{max}$ distributions are presented in Table \ref{model_table}. For example, for model 1, 50\% of the $P/P_{max}$ values are smaller than 0.0038.

\section{Results}

The variation of simulated starlight polarization across the sky, centered on the Galactic anti-center, for a subset of the S0, A0, DEHO, and axisymmetric analytic magnetic field models are shown in Figs. \ref{predic_S0} through \ref{R24_SMB}. Predictions towards the Galactic center, and in particular in the Galactic plane, are likely less reliable than the Outer Galaxy and higher Galactic latitudes because of the effects of star formation and supernovae which may interact with the large-scale magnetic field.

The S0 (Fig. \ref{predic_S0}), DEHO (Fig. \ref{predic_DEHO}), and axisymmetric analytic (Fig. \ref{predic_AA}) magnetic field models all show similar structures, with magnetic field orientations in the disk parallel to the Galactic plane and magnetic nulls near Galactic longitudes $90\degr$ and $270\degr$, which are along the toroidal magnetic component. This is significantly different from the predictions of the A0 models shown in Fig. \ref{predic_A0} where the field is predominantly perpendicular to the Galactic plane and the magnetic nulls occur at high Galactic latitudes.

Toward the $\ell=0\degr$ and $180\degr$ directions, many of the model predictions become degenerate. In particular, the three S0, three DEHO, and $\theta=0\degr$ analytic models make identical predictions toward these directions at all Galactic latitudes. The other axisymmetric analytic models vary with Galactic latitude because of projection effects alone. The $\ell=90\degr$ and $270\degr$ directions are also degenerate in all models except the $\theta=11.5\degr, 24\degr$ models (Models 21-24). Toward these directions, the field lies predominantly along the line of sight, so the toroidal component has a small projection onto the sky and the (mostly vertical) poloidal field projection dominates. This is also reflected in the weak polarizations predicted for these directions. In the models with large pitch angles ($\theta=11.5\degr, 24\degr$; Models 21-24), the magnetic nulls shift in Galactic longitude (see Fig. \ref{predic_AA}), due to the change in the magnetic field geometry. The amplitude of the longitude shift depends directly on the pitch angle. The A0 models are almost entirely perpendicular to the Galactic plane for $|b| < 45\degr$. For $|b| \geq 45\degr$, the projection of the radial magnetic field component begins to align with the Galactic plane. Because of all these effects, the directions near $\ell = 0\degr,\, 90\degr,\, 180\degr,$ and $270\degr$ are not suitable for distinguishing among the various magnetic field model geometries.

The simulated data were extracted from all the models at one example Galactic longitude ($\ell=150\degr$). As shown in Fig. \ref{S0_cut}, the S0 models are seen to all have similar shapes in the (PA, b) plane, especially when compared to the very different A0 models shown in Fig. \ref{A0_cut}. The DEHO models of \citet{M10} in Fig. \ref{DEHO_cut} have similar shapes to the S0 models of \citet{FS2000}. This is expected since the \citet{M10} DEHO models have even magnetic fields in the disk and odd magnetic fields in the halo. Both dust density models fall exponentially with Galactic height, so the modeled polarizations are more sensitive to fields in and near to the disk and less sensitive to far halo (high latitude) fields. The higher PA dispersions seen toward Galactic mid-latitudes in the DEHO models may be caused by partial sampling of the halo-odd field. The $\theta = 0\degr$ models (Models 19 and 20 in Fig. \ref{AA_cut}) are qualitatively similar to the S0 and DEHO models, however the other axisymmetric analytic models (in Fig. \ref{AA_cut}) show a flattening of their PA vs. b for the higher magnetic pitch angles. By $\theta = 24\degr$ (Models 23 and 24 in Fig \ref{AA_cut}), the distribution of PA as a function of Galactic latitude is nearly flat, with mean PAs around $90\degr$. In this case, the degenerate region that exists near $\ell=180\degr$  for $\theta=0\degr$ has shifted towards $\ell=150\degr$ (as seen in Fig. \ref{R24_DS}).

The degeneracies typically seen towards $\ell=90\degr$ and $270\degr$ for the even models with small magnetic pitch angles (Models 1-6, 19, and 20) was broken for larger pitch angles (Models 21-24), thus the longitude location of the polarization nulls may be used as a tool for measuring the magnetic pitch angle. \citet{H96} has previously used the location of these polarization null points to estimate the Galactic magnetic pitch angle. Towards $\ell=90\degr$, projections of the magnetic pitch angle causes a distinctive change in the shape of the PA vs. Galactic latitude plot. This may allow observations in this direction to put constraints on the magnetic pitch angle of the Milky Way. It is important to note however, that CDFs for this region of the sky were dominated by low polarizations, and that many of the predicted stellar polarizations may fall below realistic detection thresholds.

The choice of dust models has a non-negligible effect. As an example, consider models 13 and 14 in Fig. \ref{DEHO_cut}; both models use the same model stars, the same model magnetic field from \citet{M10}, but use different dust models (\citealt{SMB96} versus \citealt{DS01}). The RMS variation of the mean PA values across the sky between the two models is $1.09\degr$, which is small or comparable to the mean PA dispersion for each model ($0.91\degr$ for Model 13, and $1.77\degr$ for Model 14). The most significant difference between the two models is found in the intrinsic PA dispersions, shown as a ratio in Fig. \ref{dust_example} for the one Galactic longitude $\ell=150\degr$. The differences seen must be due to the dust model details. This effect was seen along all simulated Galactic longitudes. The \citet{DS01} dust model (used for Model 14) includes the local arm, spiral arms, and a Galactic warp, which were not included in the \citet{SMB96} model (used for Model 13). The choice of dust model also affects the tabulated CDF vales for the different models, as listed in Table \ref{model_table}. For any of the four classes of magnetic fields used, there are differences among the three \citet{DS01} models and the three \citet{SMB96} models. For the A0, S0, and axisymmetric analytic magnetic fields, the \citet{DS01} models produce slightly higher normalized polarizations than the \citet{SMB96} models. DEHO magnetic fields show the opposite effect, with the \citet{DS01} models producing slightly lower normalized polarizations than \citet{SMB96} based models.

All of these simulations are optically thin, and so wavelength independent. However, wavelength is known to affect the polarizing efficiency of dust grains \citep{SMF75}. Under the assumption that there is a uniform wavelength dependence of polarizing efficiency along the lines of sight, scaling between wavelengths only requires use of a multiplicative factor between the predicted, normalized degree of polarization distribution and the actual polarization distribution. These effects will not change the observed PA for a given star since, under these assumptions, the polarizing efficiency does not affect the PA calculation. 

\begin{figure}
	\includegraphics[scale=0.92]{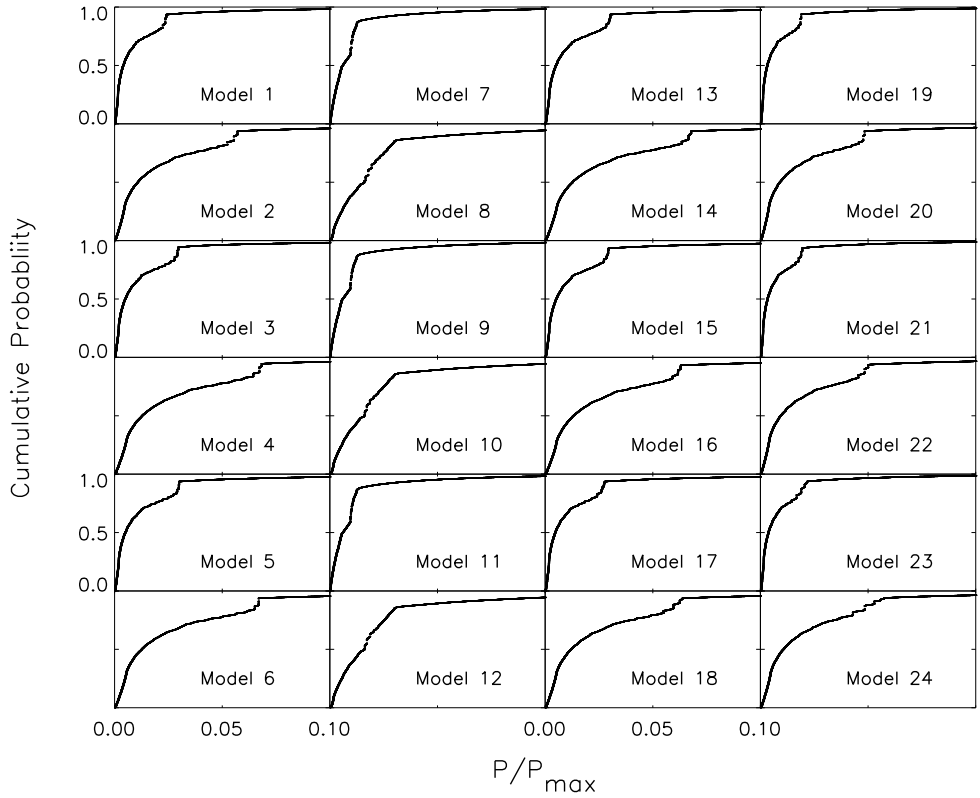}
	\caption{\label{CDFs}Cumulative probability distributions (CDFs) of stellar polarization for all 24 models, normalized by the maximum predicted degree of polarization, shown for cumulative probabilities of 0 to 0.1. These CDFs can be used to calibrate the predicted degree of starlight polarization with real observations.}
\end{figure}

In Fig. \ref{CDFs}, the CDFs of the P values in each model are shown, normalized by the star with the highest predicted polarization. Each model was normalized independently. The star with the highest predicted polarization value is typically 20-40 times the average modeled polarization. In 75\% of the models, the star with the highest polarization was a simulated G2 supergiant at 13.31 kpc, and in 25\% of the models it was a simulated F8 supergiant at 13.51 kpc (all of the models used the same stellar population samples). These were some of the most distant stars returned by the \citet{R03} stellar population model. The long distances between these stars and observer generated larger degrees of starlight polarization. These bright stars are exceedingly rare, but may be detected in magnitude-limited polarization surveys covering large fractions of the sky. Since these high polarization stars make up such a small fraction of the total population, the CDFs shown in Fig. \ref{CDFs} only span the lowest ten percent of the entire CDF.

A discontinuity is seen in the CDFs at approximately the 93rd percentile, where distant, high polarization stars occur. There are also differences to be noted between the odd-numbered models, which used the \citet{DS01} dust distribution, and the even-numbered models, which use the \citet{SMB96} dust distribution. For a given magnetic field geometry, the models using the \citet{SMB96} dust distribution show predicted polarizations typically about four times larger than the models using the \citet{DS01} dust distribution. This effect is strongest for low polarization stars and weaker for the stars with the highest polarizations. Normalization by the maximum predicted polarization does not mask the effect. It is caused by differences in the normalization constants used for the two dust distributions. The dust column was calculated for every sample star using the two different dust distributions. The average dust column ratio was 4.2, which is approximately the maximum value in a corresponding polarization ratio distribution. The interplay between the exact dust distribution details and projected magnetic field geometry creates some depolarization, which causes the polarization ratio to decrease for some stars.

In addition to an unknown polarization scaling factor, actual observations will suffer from censoring at the low P end of the CDF, as stars fall below polarization detection thresholds. Stars above detection thresholds can be used to calculate an appropriate scaling factor to match observations to the current simulated predictions. In practice, the censoring of P below an observational pfolarization limit would manifest as a sharp cutoff at the low polarization end of the CDF. The low and high polarization ends of the CDF may also suffer from undersampling. The 25th, 50th, and 75th percentile CDF polarization values should therefore best serve as appropriate locations to calculate this scaling factor, and these values are tabulated for each of the model CDFs and presented in Table \ref{model_table}.

In a magnitude-limited polarimetric survey, the observed polarization CDF of all stars (including non-detections) above the observational apparent magnitude limit would be constructed. A predicted degree of polarization CDF in the same region of the sky, at the same limiting magnitude, should be simulated. The ratios of the degrees of polarization at a given percentile value in the observed and simulated CDFs will give the normalization factor. This normalization factor can then be used to calibrate the predicted degrees of starlight polarization to actual observations.

\section{Discussion}

The simulations presented here attempt to bridge the gap between theory and observation by making testable predictions for starlight polarization observations based on models of the large-scale structure of the Galactic magnetic field. These predictions, when combined with NIR polarization observations, can be used to test the models for the large-scale magnetic field and their physical underpinnings. Polarization of background starlight is insensitive to magnetic reversals and cannot address the presence or location of reversals. However, these polarizations can allow testing for the existence and nature of the poloidal component of the Galactic magnetic field, since all S0 models yield similar polarization predictions across the sky and all A0 models yield similar predictions that are distinctly different from the S0 models. The predicted longitude shift of the polarization null points with magnetic pitch angle, demonstrated in Models 19-24 (a small shift is seen all of the models, however only Models 19-24 systematically vary the pitch angle), may facilitate an observational test for the Galactic magnetic pitch angle in the disk, similar to the work of \citet{H96}. This test is fundamentally different from previous attempts to use Faraday rotation of pulsars and extragalactic sources to model the magnetic pitch angle \citep{RK89,MFH08,NK10,VE11}.

The Galaxy was assumed to have an axisymmetric magnetic field based on theory \citep{Ruz88,B90,MB92} and observational evidence \citep{RK89,RL94,Sun08,R10}, but there is some evidence in other galaxies \citep[particularly in M81,][]{K89,S92} that bisymmetric fields may exist. \citet{B96} indicate that many of the galaxies showing evidence for bisymmetric magnetic fields also show evidence for galaxy interactions, so we might not expect this magnetic field structure for the Milky Way (based on a lack of major merger events; \citealt{GWN02}) and the axisymmetric assumption may hold. Higher order azimuthal symmetries might also be possible, but their amplitudes should be relatively small \citep{B96}.

The measurements needed to distinguish among the various model simulations are well suited to NIR stellar polarimetry. The polarization mechanism used in the simulations works from the NIR through near-UV wavelengths \citep{SMF75,CM76,W92}. However, NIR light is less attenuated by interstellar dust and can probe magnetic fields along multi-kpc scales, while the shorter wavelengths only probe within about 1 kpc \citep{F02}. As described by \citet{SMF75}, the polarization signal in the NIR is weaker than the visible by a factor of four or more. However, NIR polarimetric observations at this level are possible with recent instrumentation \citep{K06,C07} and should soon provide data able to test models of the large-scale structure of the Galactic magnetic field.

The simplest comparison may be through collecting the polarization behavior exhibited through a set of samples of all Galactic latitudes at a single Galactic longitude, as shown in Figs. \ref{S0_cut} through \ref{AA_cut}. To best understand the poloidal component of the Galactic magnetic field, these observations should be made in the outer Galaxy longitude ranges of $\ell=110-160\degr$ or $\ell=200-250\degr$. These ranges avoid the degenerate regions that reside near $\ell=0,\,90,\,180$, and $270\degr$ with a $20\degr$ buffer zone in either direction to allow for possible wander caused by any reasonable magnetic pitch angles (if present). To measure this magnetic pitch angle, a sample of Galactic longitudes should be observed at a single Galactic latitude, say $b=15\degr$, similar to the approach of \citet{H96}.

\section{Summary}

Predictions for the polarization of background starlight, based on physics-driven Galactic dynamo simulations and empirical Galactic dust distributions, are presented. A Stokes radiative transfer model was used to predict the observed stellar polarization properties from Monte Carlo generated realistic stellar distributions in the Galaxy. A range of sky directions are suggested as particularly diagnostic regions for differentiating among Galactic magnetic field model geometries. Samples across many Galactic longitudes at one Galactic latitude, however, may be more appropriate for characterizing the magnetic pitch angle of magnetic field patterns in the Galaxy.

The CDF of observed starlight polarizations is proposed as a tool for calibrating these models. Table \ref{model_table} presents quartile values, for each model, of the normalized degree of starlight polarization. These measures are easily obtained from observations of starlight polarizations and can be used to determine a scaling coefficient between the predicted degree of polarization and observed degree of polarization.

The role of the Galactic magnetic field in interstellar dynamics and star formation is only beginning to be understood, but new technical advances will permit probing this mysterious component of our Galaxy. The study of the Galactic magnetic field has been dominated by Faraday rotation measurements, but new tools, such as NIR polarization of background starlight, complement these studies.

\acknowledgements
The author thanks D. Schmitt and D. Moss for providing electronic versions of their simulations, and K. Ferri\`ere, T. J. Jones, D. Clemens, and A. Pinnick for helpful comments and discussions. The author would also like to thank the anonymous referee whose comments significantly improved this work. This work was partially supported by NSF grants AST 06-07500 and 09-07790 to Boston University, D. Clemens PI.

\begin{deluxetable}{cccccccc}
	\tabletypesize{\footnotesize}
	\tablecaption{Model Designations \label{model_table}}
	\tablewidth{0pt}
	\rotate
	\tablecolumns{6}
	\tablehead{\multirow{2}{*}{Model Number} &
						 \multirow{2}{*}{Magnetic Field} &
						 \colhead{Magnetic} &
						 \colhead{Dust Model} &
						 \multicolumn{4}{c}{\underline{  P/$P_{max}$ Percentiles [$\times 10^{4}$] }} \\
						 \colhead{} & 
						 \colhead{} & 
						 \colhead{Pitch Angle} &
						 \colhead{Reference\tablenotemark{a}} & 
						 \colhead{25th} & 
						 \colhead{} &
						 \colhead{50th} & 
						 \colhead{75th} \\
						 }
	\startdata
			1 & S0 reference run\tablenotemark{b} & 8.4\degr & DS2001 & 13.6 & & 38.3 & 151 \\
			2 & S0 reference run\tablenotemark{b} & 8.4\degr & SMB96  & 40.2 & & 106 & 367 \\
			3 & S0 reference with $\alpha$ quenching\tablenotemark{b} & 9.1\degr & DS2001 & 16.8 & & 47.6 & 178 \\
			4 & S0 reference with $\alpha$ quenching\tablenotemark{b} & 9.1\degr & SMB96  & 48.1 & & 126 & 417 \\
			5 & S0 with vacuum BC\tablenotemark{b} & 9.1\degr & DS2001  & 16.8 & & 47.8 & 189 \\
			6 & S0 with vacuum BC\tablenotemark{b} & 9.1\degr & SMB96  & 47.0 & & 124 & 427 \\
			\hline
			7 & A0 reference run\tablenotemark{b} & 10.6\degr & DS2001  & 21.1 & & 57.8 & 103 \\
			8 & A0 reference run\tablenotemark{b} & 10.6\degr & SMB96  & 53.5 & & 157 & 248 \\
			9 & A0 reference with $\alpha$ quenching\tablenotemark{b} & 7.0\degr & DS2001 & 21.3 & & 57.7 & 104 \\
			10 & A0 reference with $\alpha$ quenching\tablenotemark{b} & 7.0\degr & SMB96 & 54.0 & & 157 & 249 \\
			11 & A0 with vacuum BC\tablenotemark{b} & 7.3\degr & DS2001 & 21.3 & & 58.1 & 104 \\
			12 & A0 with vacuum BC\tablenotemark{b} & 7.3\degr & SMB96 & 54.0 & & 158 & 251 \\
			\hline
			13 & $C_{wind} = 0, R_{\alpha,halo} = 300$\tablenotemark{c} & 8.1\degr & DS2001 & 17.0 & & 48.3 & 188 \\
			14 & $C_{wind} = 0, R_{\alpha,halo} = 300$\tablenotemark{c} & 8.1\degr & SMB96 & 47.6 & & 125 & 429 \\
			15 & $C_{wind} = 100, R_{\alpha,halo} = 300$\tablenotemark{c} & 13.0\degr & DS2001 & 17.4 & & 49.8 & 193 \\
			16 & $C_{wind} = 100, R_{\alpha,halo} = 300$\tablenotemark{c} & 13.0\degr & SMB96 & 48.2 & & 129 & 439 \\
			17 & $C_{wind} = 200, R_{\alpha,halo} = 300$\tablenotemark{c} & 21.0\degr & DS2001 & 15.9 & & 44.6 & 170 \\
			18 & $C_{wind} = 200, R_{\alpha,halo} = 300$\tablenotemark{c} & 21.0\degr & SMB96 & 45.1 & & 119 & 398 \\
			\hline
			19 & Analytic Model, $\theta=0\degr$ & 0\degr & DS2001 & 10.6 & & 30.0 & 112 \\
			20 & Analytic Model, $\theta=0\degr$ & 0\degr & SMB96 & 33.7 & & 88.1 & 298 \\
			21 & Analytic Model, $\theta=11.5\degr$ & 11.5\degr & DS2001 & 11.0 & & 31.4 & 120 \\
			22 & Analytic Model, $\theta=11.5\degr$ & 11.5\degr & SMB96 & 35.2 & & 91.7 & 311 \\
			23 & Analytic Model, $\theta=24\degr$ & 24\degr & DS2001 & 12.7 & & 35.0 & 125 \\
			24 & Analytic Model, $\theta=24\degr$ & 24\degr & SMB96 & 40.5 & & 105 & 339 \\
	\enddata
	\tablenotetext{a} {DS2001 = \citet{DS01};	SMB96 = \citet{SMB96}.}
	\tablenotetext{b} {\citet{FS2000}}
	\tablenotetext{c} {\citet{M10}}
\end{deluxetable}

\begin{landscape}
\begin{figure}
	\includegraphics{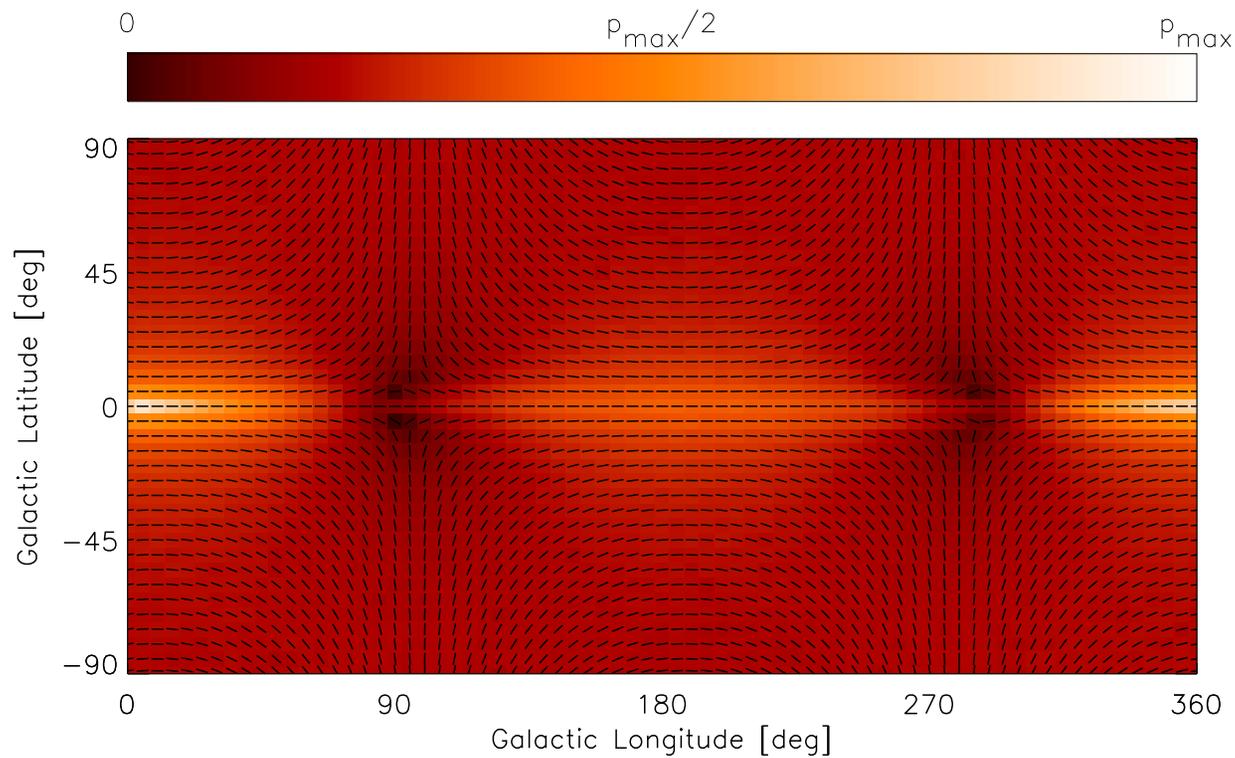}
	\centering
	\caption{\label{predic_S0}Predicted mean Galactic P and PA for $10\arcmin \times 10\arcmin$ fields on a $5\times 5\degr$ grid for the S0 Model 1. At each grid location, the uniform length vectors represent the PA of the observed polarization and the background color represents the degree of polarization. All of the S0 magnetic field Models produce a similar shape. Figs. 4-8, 10-14, 16-22, 24, and 26 are available in the online version of the Journal.}
\end{figure}

\begin{figure}
	\includegraphics{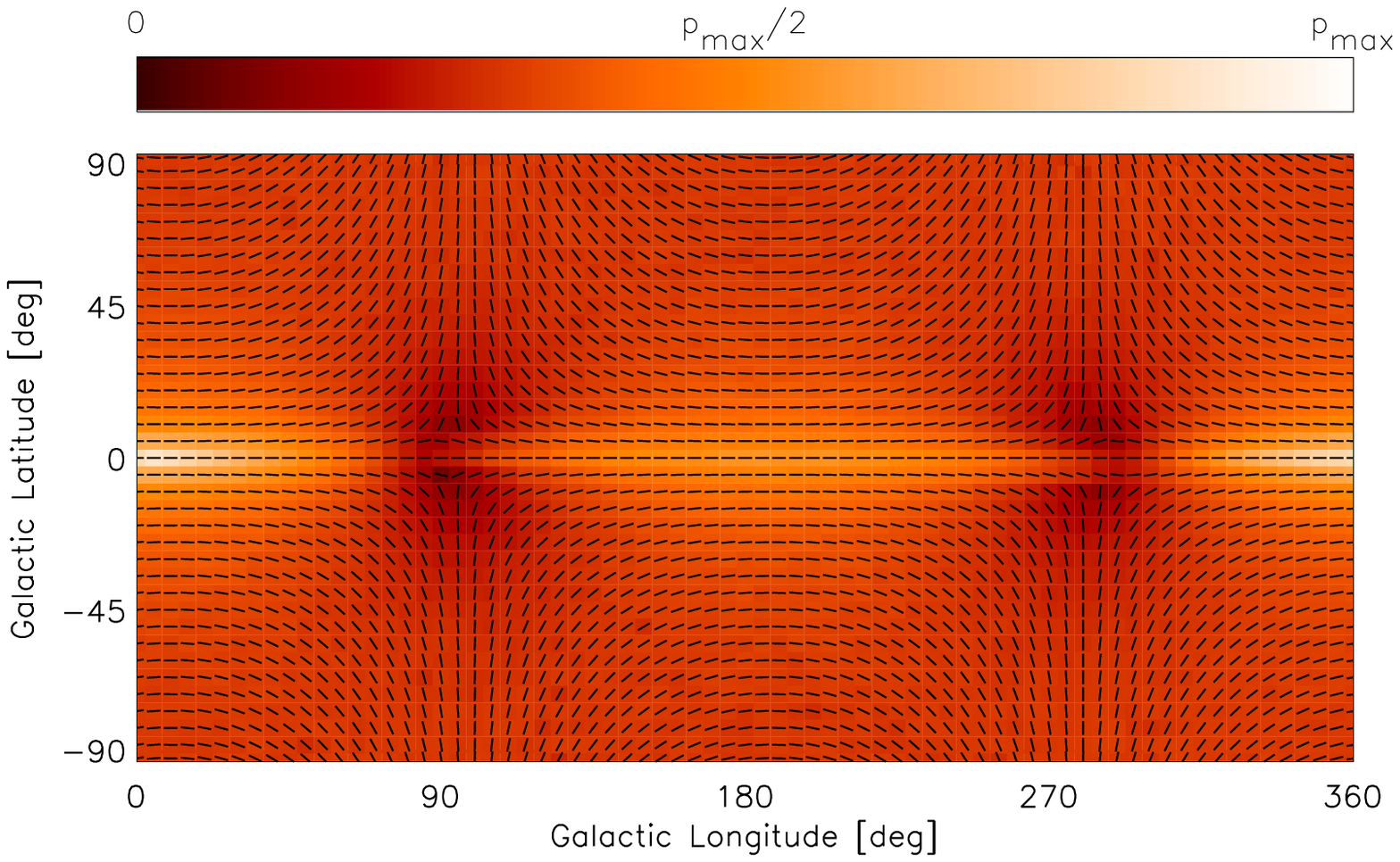}
	\centering
	\caption{\label{FS2_SMB}Same as Fig. \ref{predic_S0}, but for S0-type Model 2.}
\end{figure}

\begin{figure}
	\includegraphics{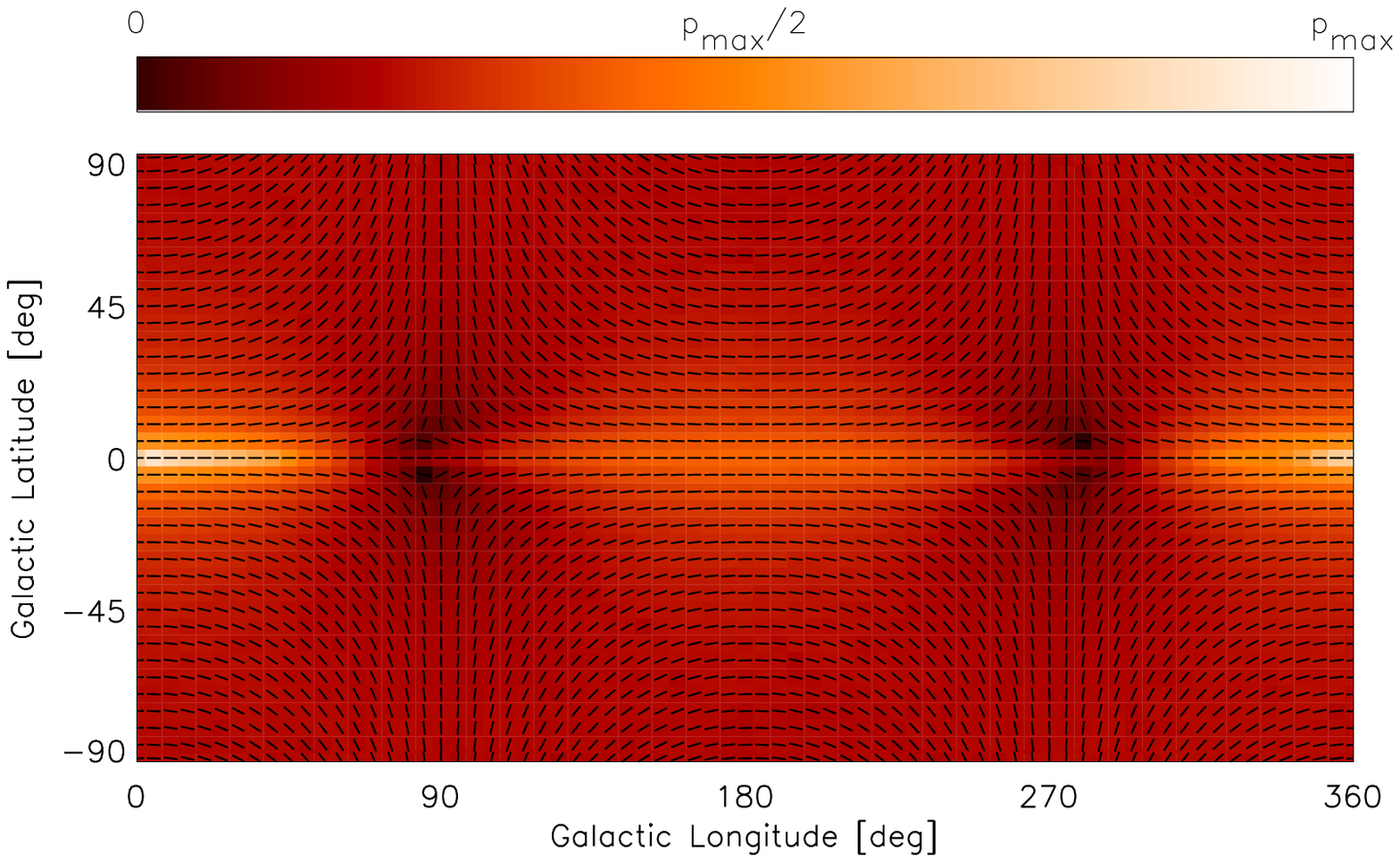}
	\centering
	\caption{\label{FS3_DS}Same as Fig. \ref{predic_S0}, but for S0-type Model 3.}
\end{figure}

\begin{figure}
	\includegraphics{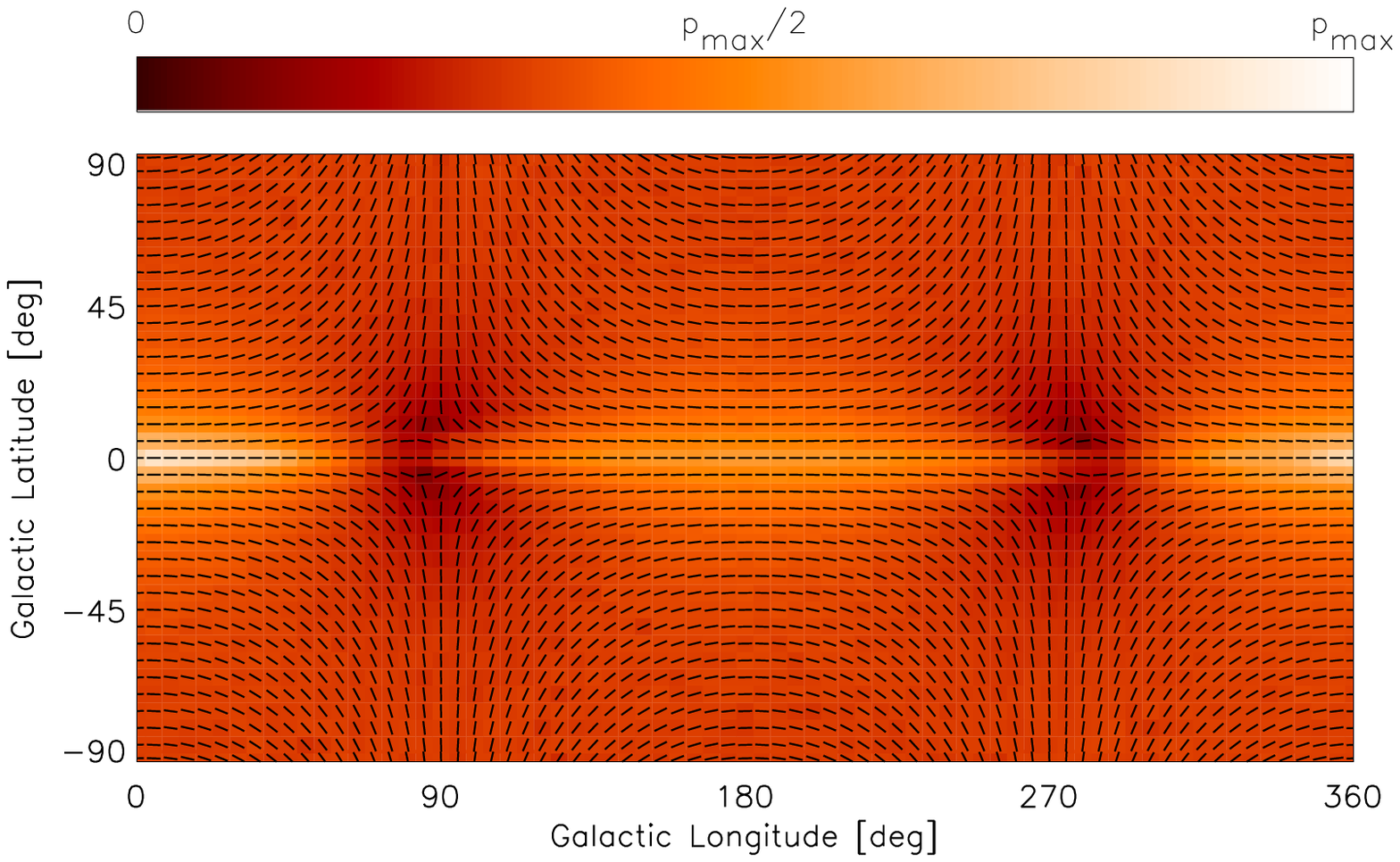}
	\centering
	\caption{\label{FS3_SMB}Same as Fig. \ref{predic_S0}, but for S0-type Model 4.}
\end{figure}

\begin{figure}
	\includegraphics{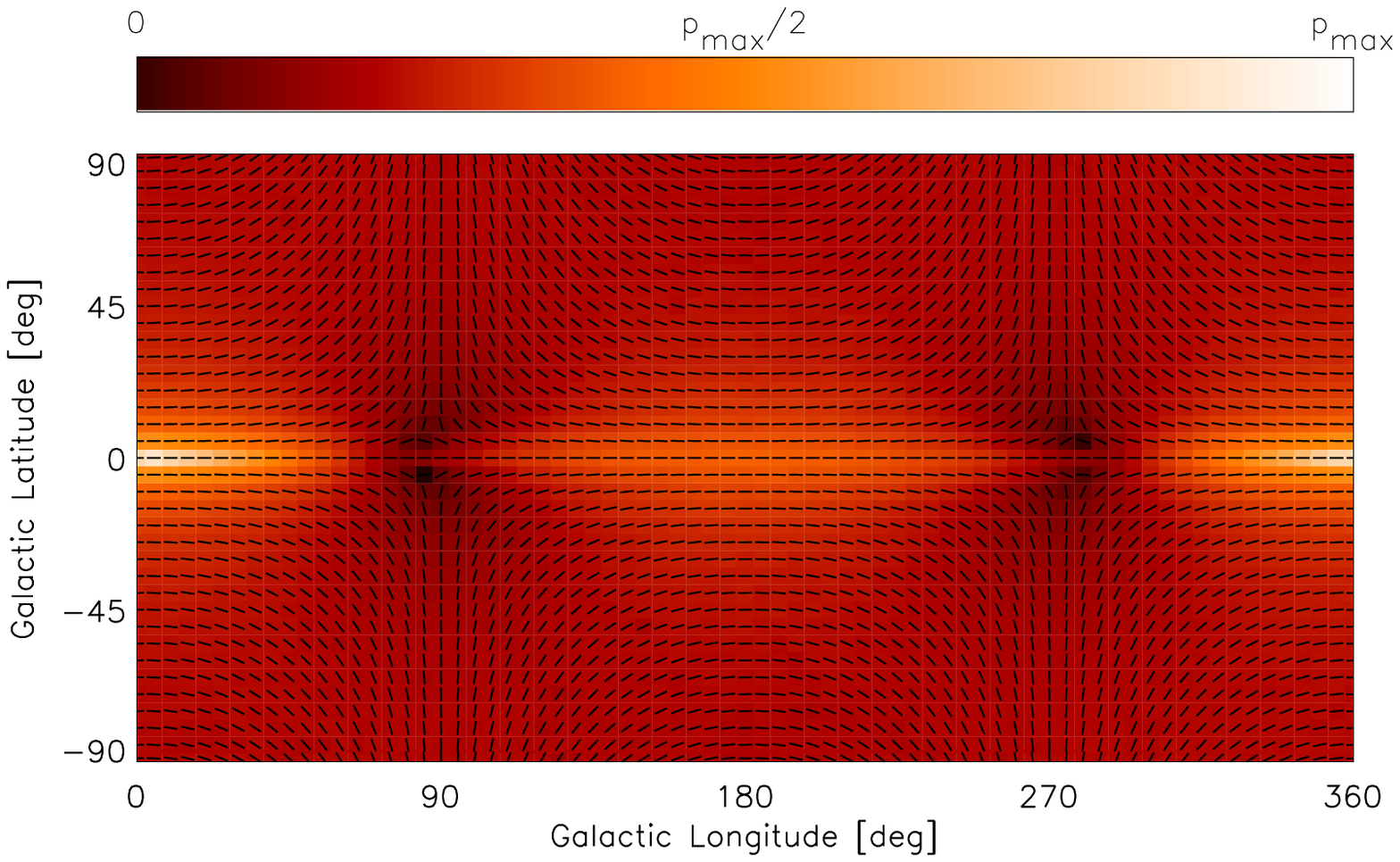}
	\centering
	\caption{\label{FS4_DS}Same as Fig. \ref{predic_S0}, but for S0-type Model 5.}
\end{figure}

\begin{figure}
	\includegraphics{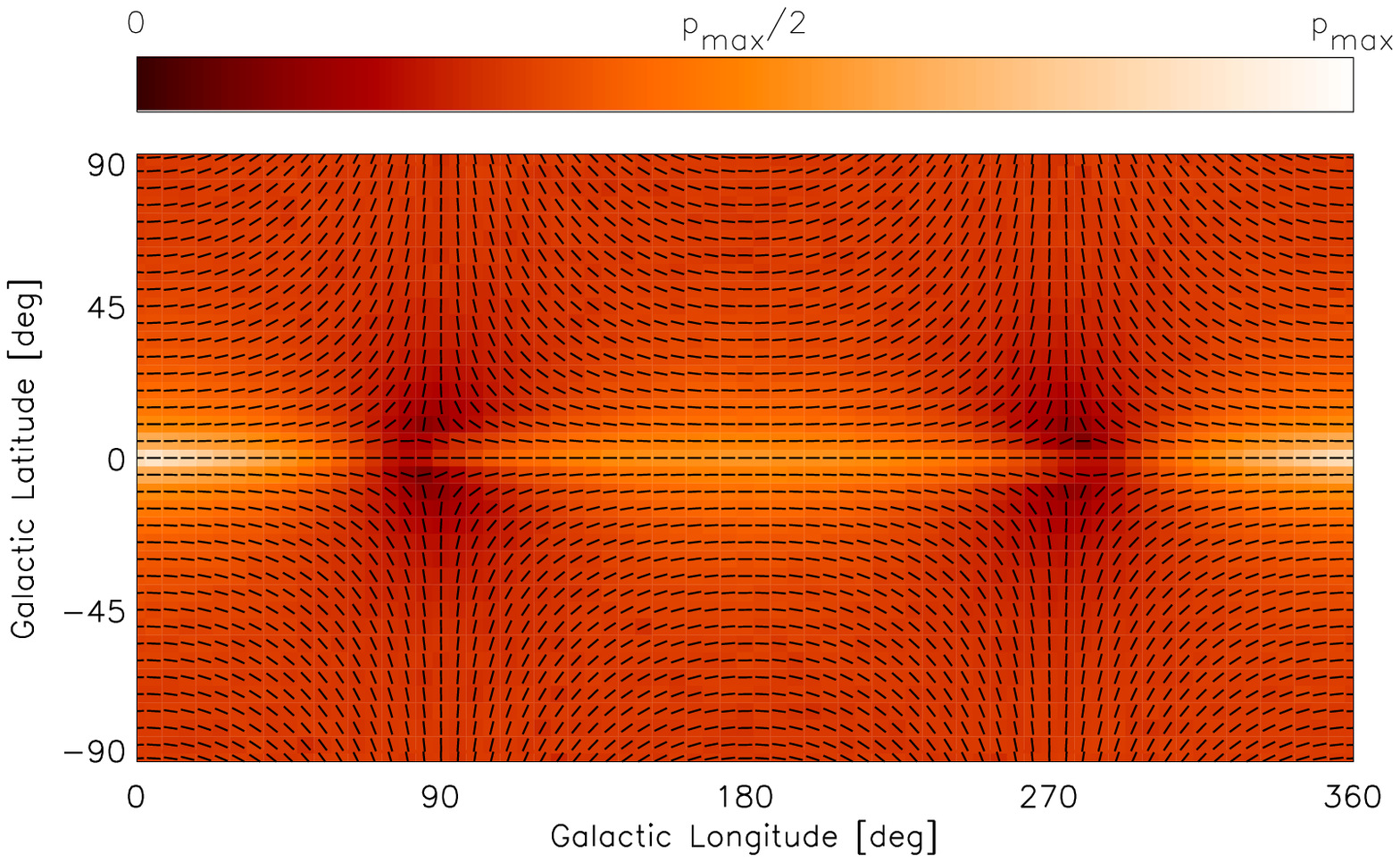}
	\centering
	\caption{\label{FS4_SMB}Same as Fig. \ref{predic_S0}, but for S0-type Model 6.}
\end{figure}

\begin{figure}
	\includegraphics{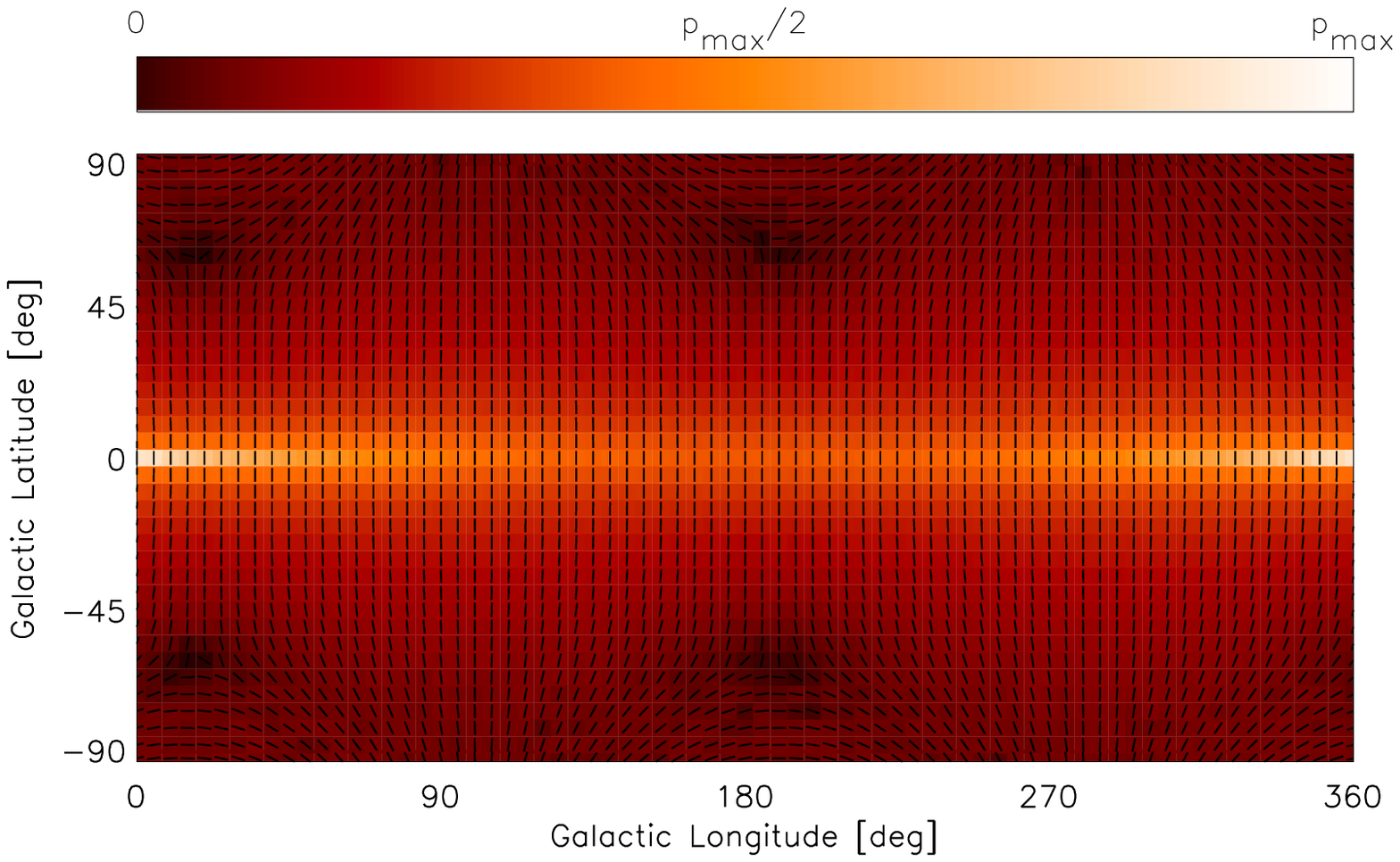}
	\centering
	\caption{\label{predic_A0}Same as Fig. \ref{predic_S0}, but for the A0-type Model 7.}
\end{figure}

\begin{figure}
	\includegraphics{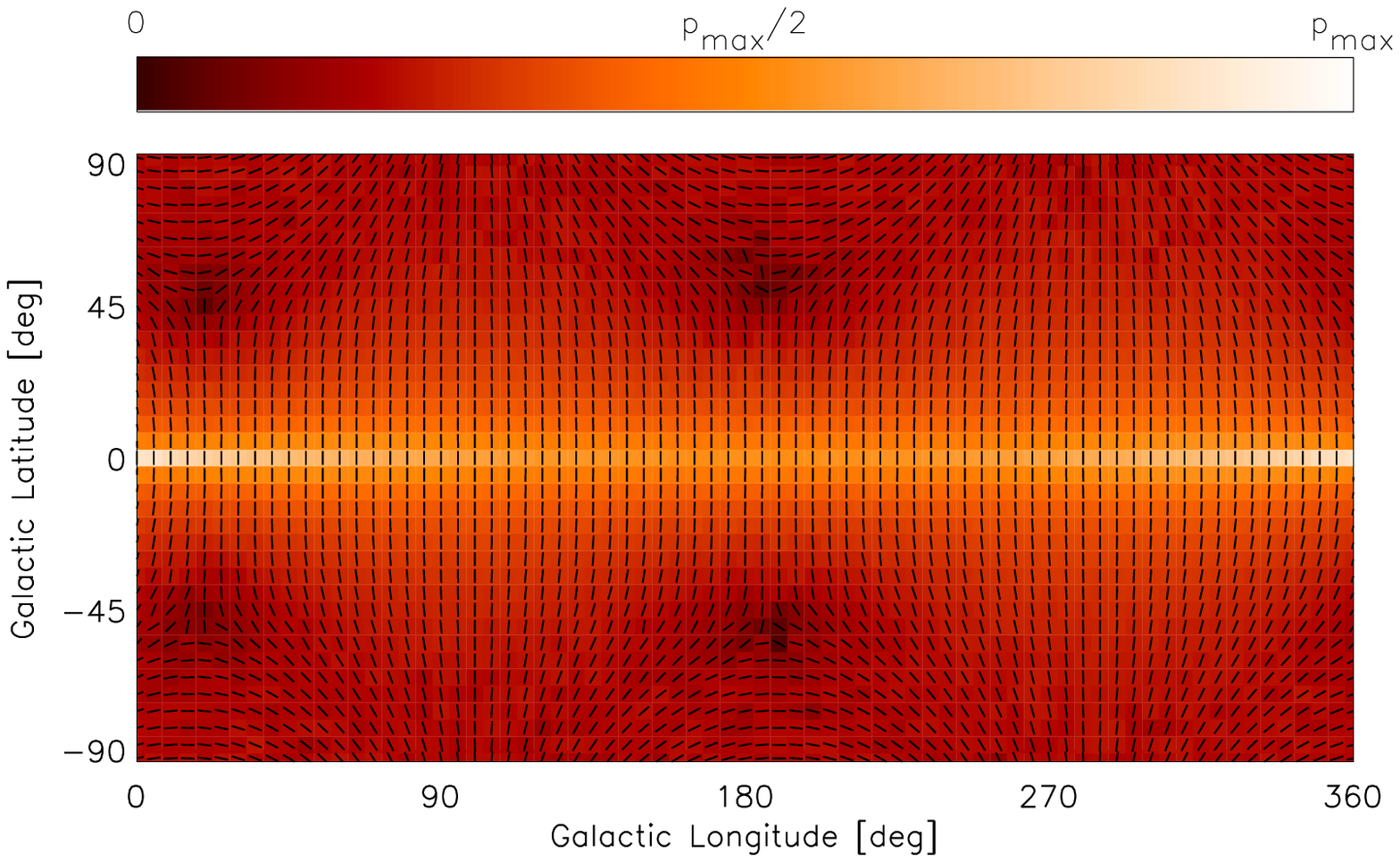}
	\centering
	\caption{\label{FS7_SMB}Same as Fig. \ref{predic_S0}, but for the A0-type Model 8.}
\end{figure}

\begin{figure}
	\includegraphics{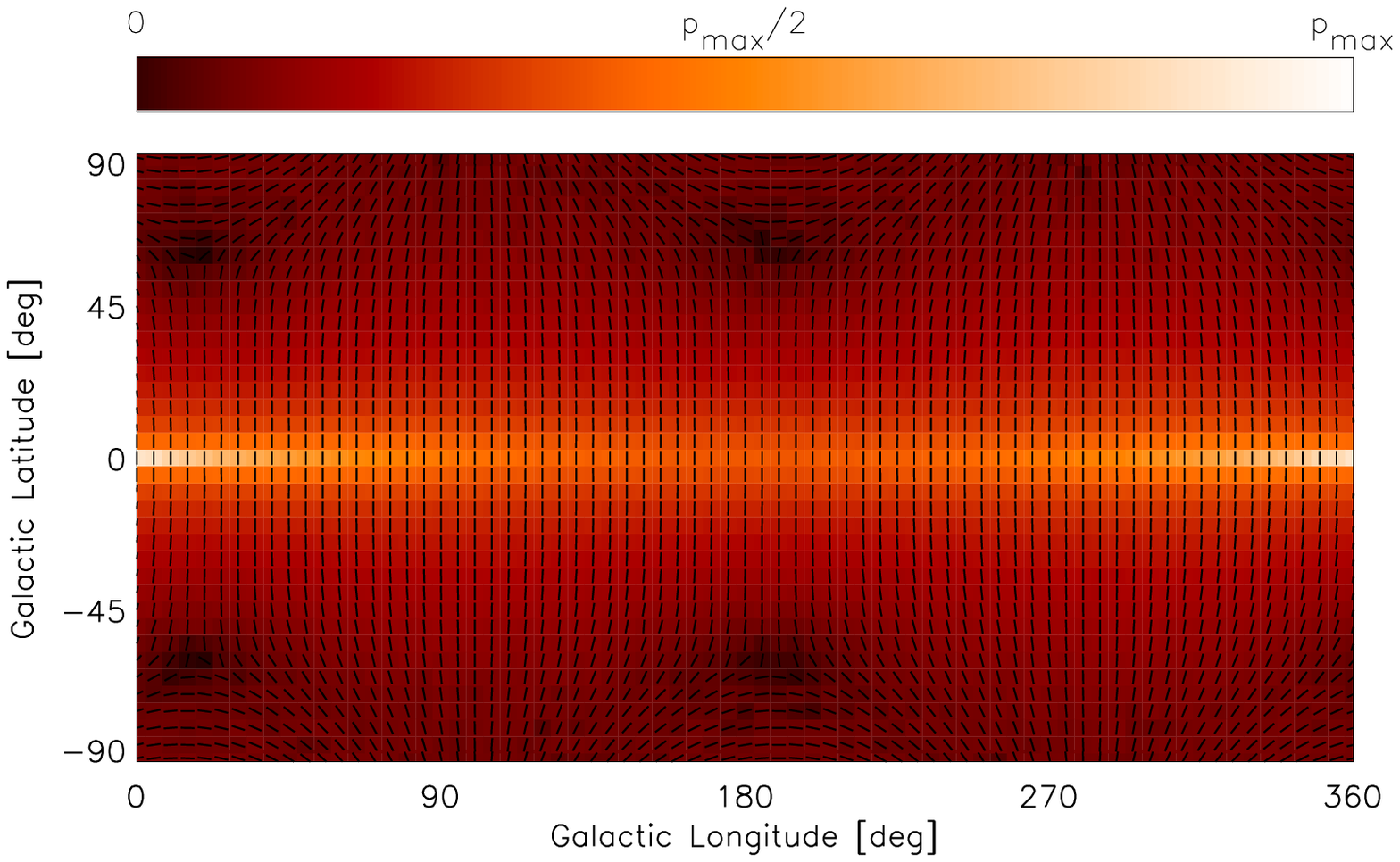}
	\centering
	\caption{\label{FS8_DS}Same as Fig. \ref{predic_S0}, but for the A0-type Model 9.}
\end{figure}

\begin{figure}
	\includegraphics{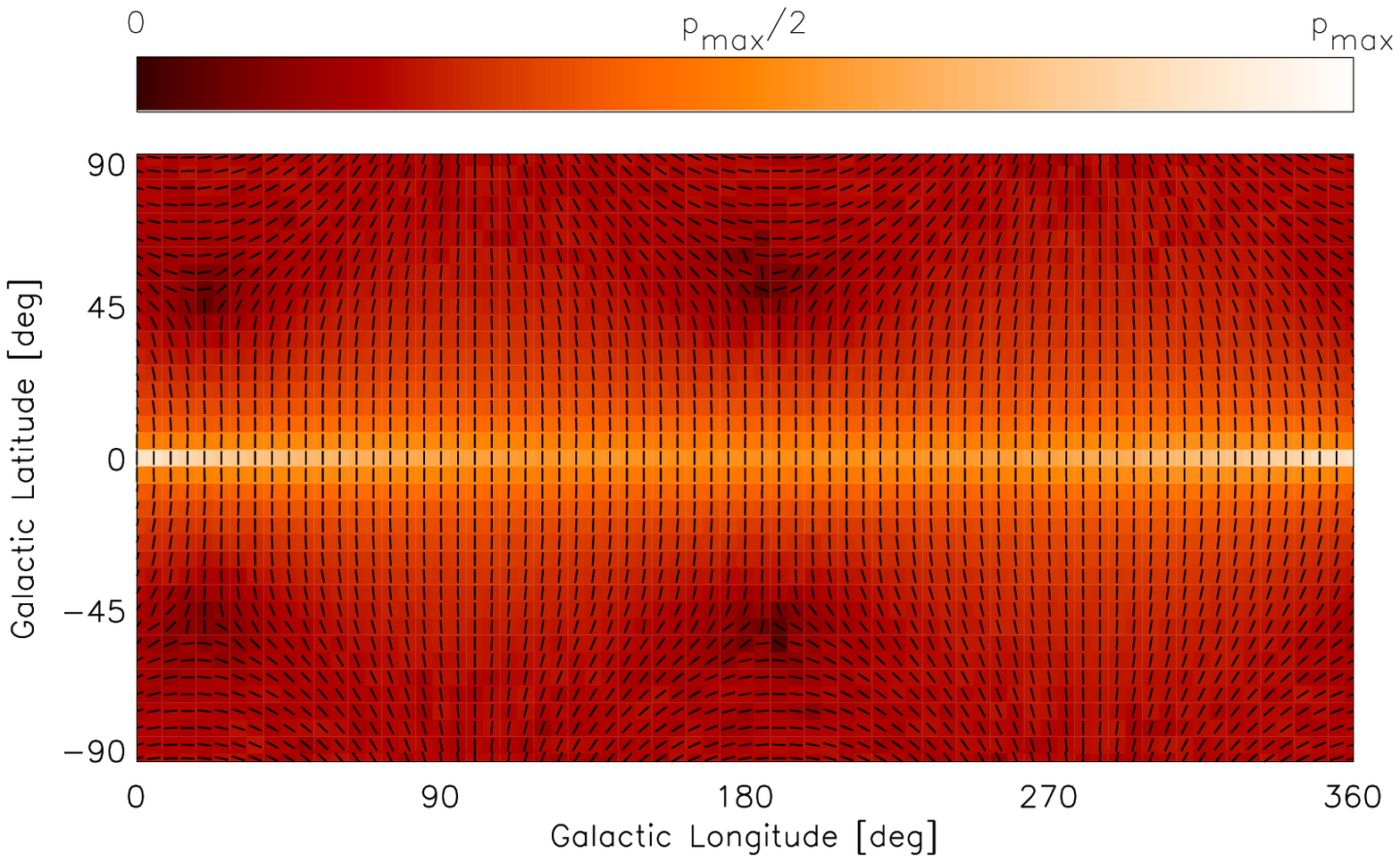}
	\centering
	\caption{\label{FS8_SMB}Same as Fig. \ref{predic_S0}, but for the A0-type Model 10.}
\end{figure}

\begin{figure}
	\includegraphics{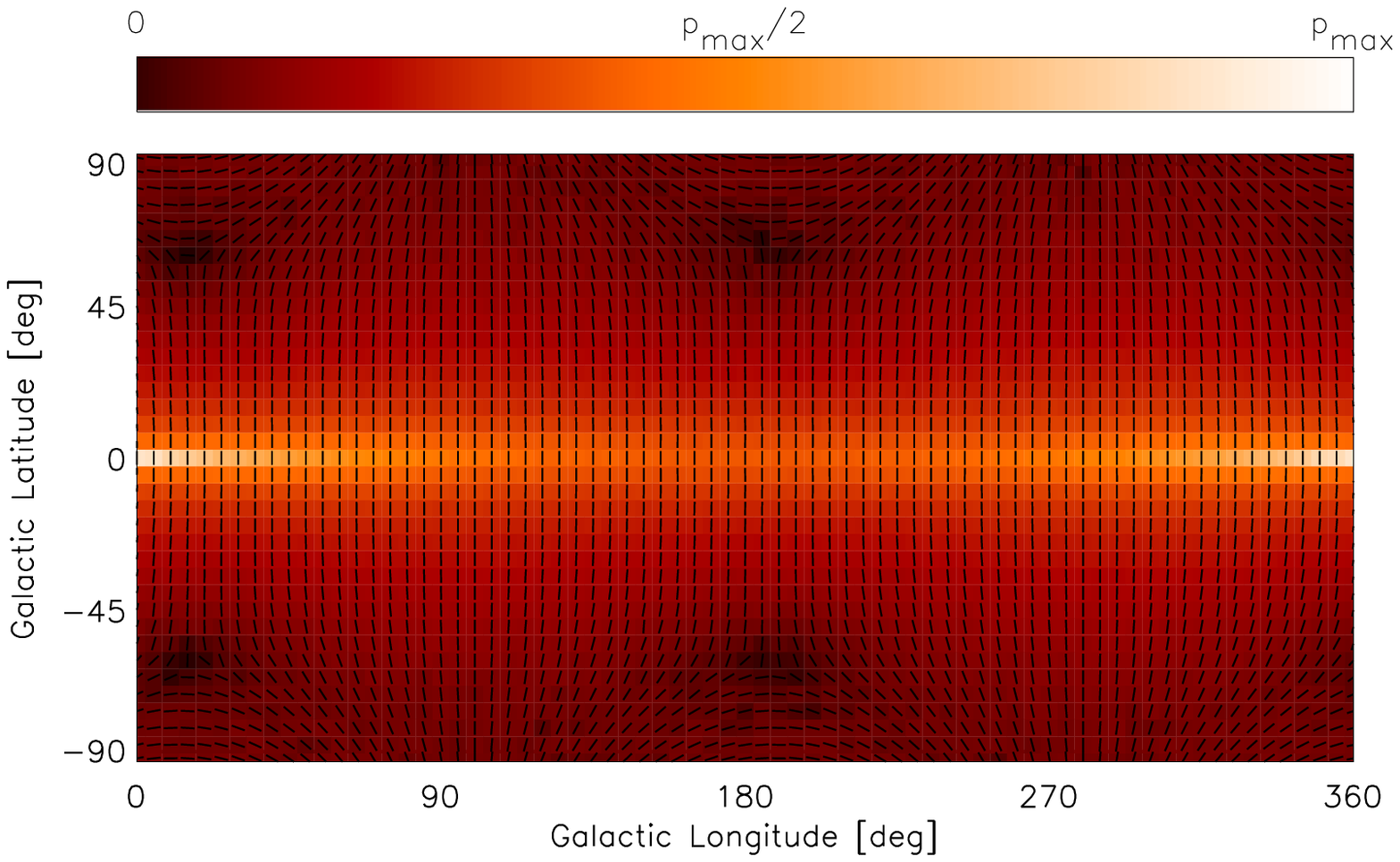}
	\centering
	\caption{\label{FS9_DS}Same as Fig. \ref{predic_S0}, but for the A0-type Model 11.}
\end{figure}

\begin{figure}
	\includegraphics{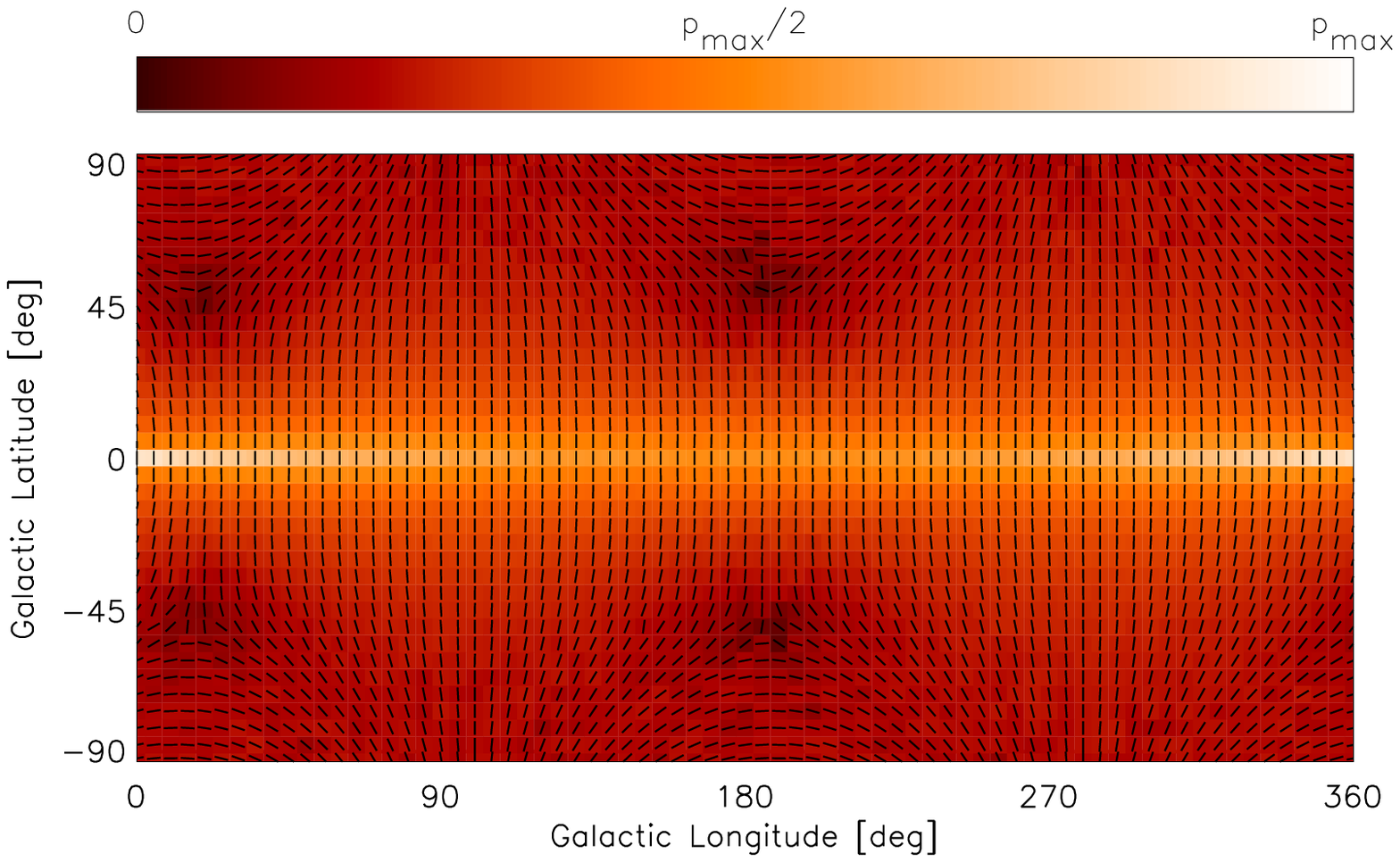}
	\centering
	\caption{\label{FS9_SMB}Same as Fig. \ref{predic_S0}, but for the A0-type Model 12.}
\end{figure}

\begin{figure}
	\includegraphics{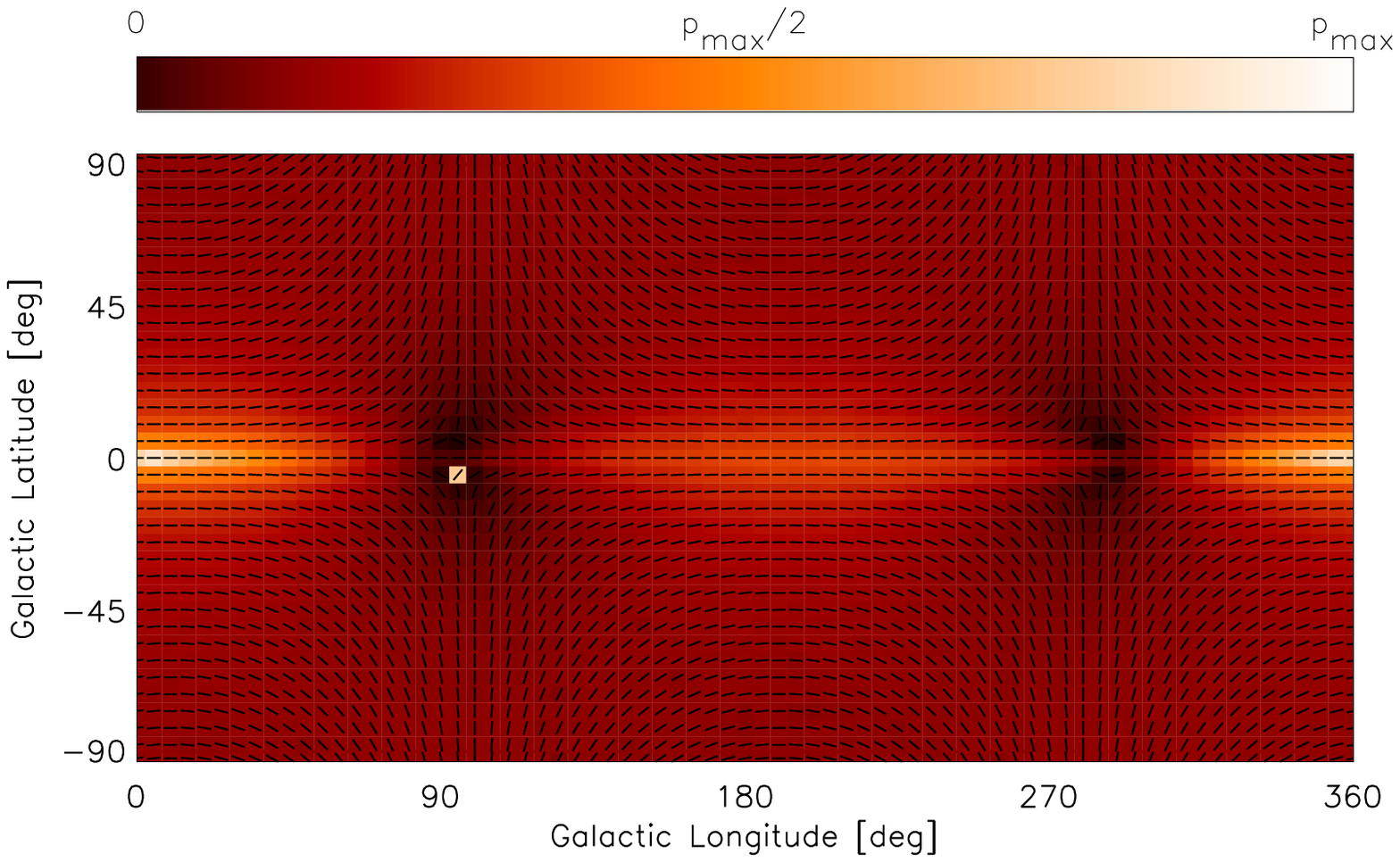}
	\centering
	\caption{\label{predic_DEHO}Same as Fig. \ref{predic_S0}, but for the DEHO Model 13.}
\end{figure}

\begin{figure}
	\includegraphics{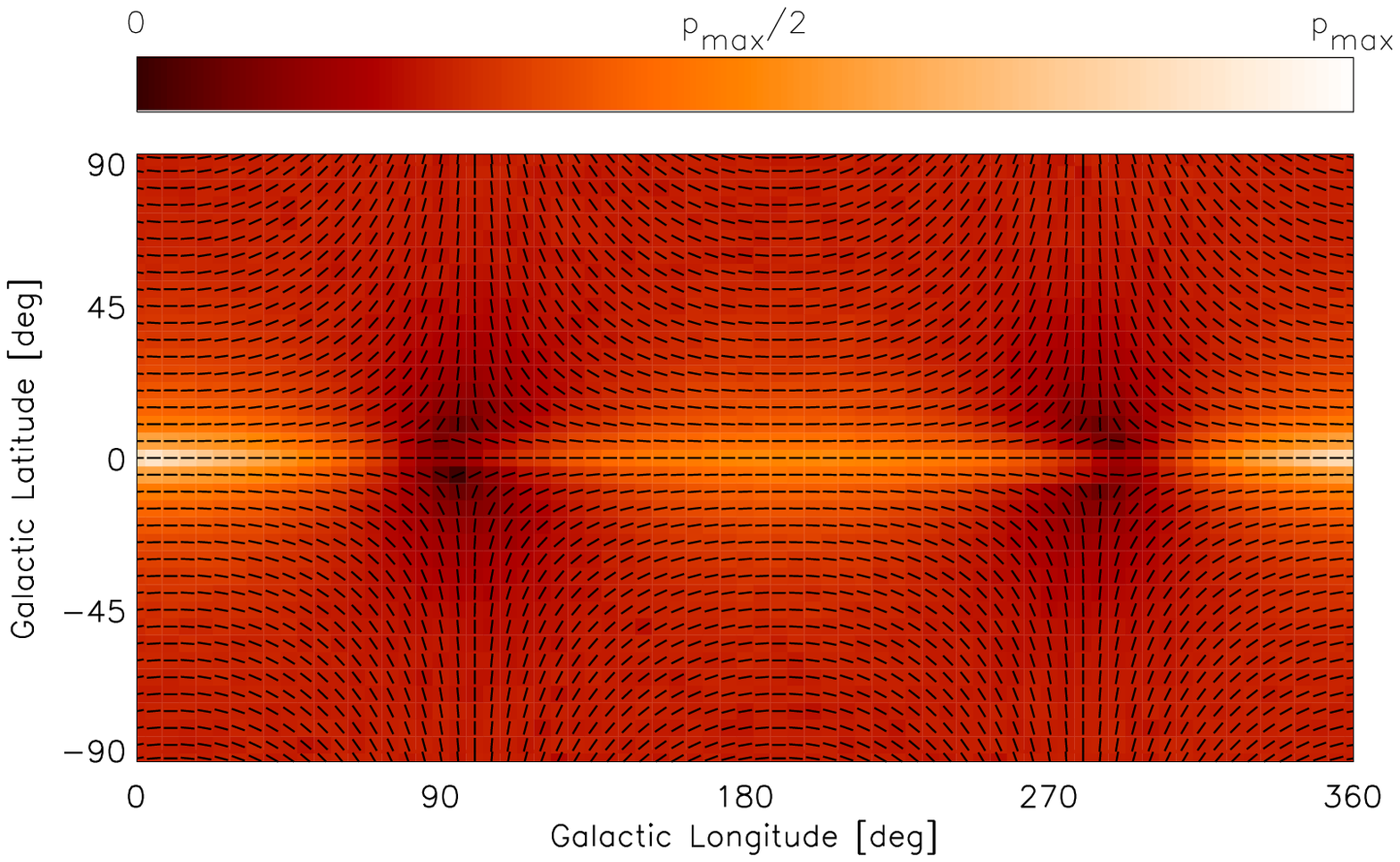}
	\centering
	\caption{\label{M511_SMB}Same as Fig. \ref{predic_S0}, but for the DEHO Model 14.}
\end{figure}

\begin{figure}
	\includegraphics{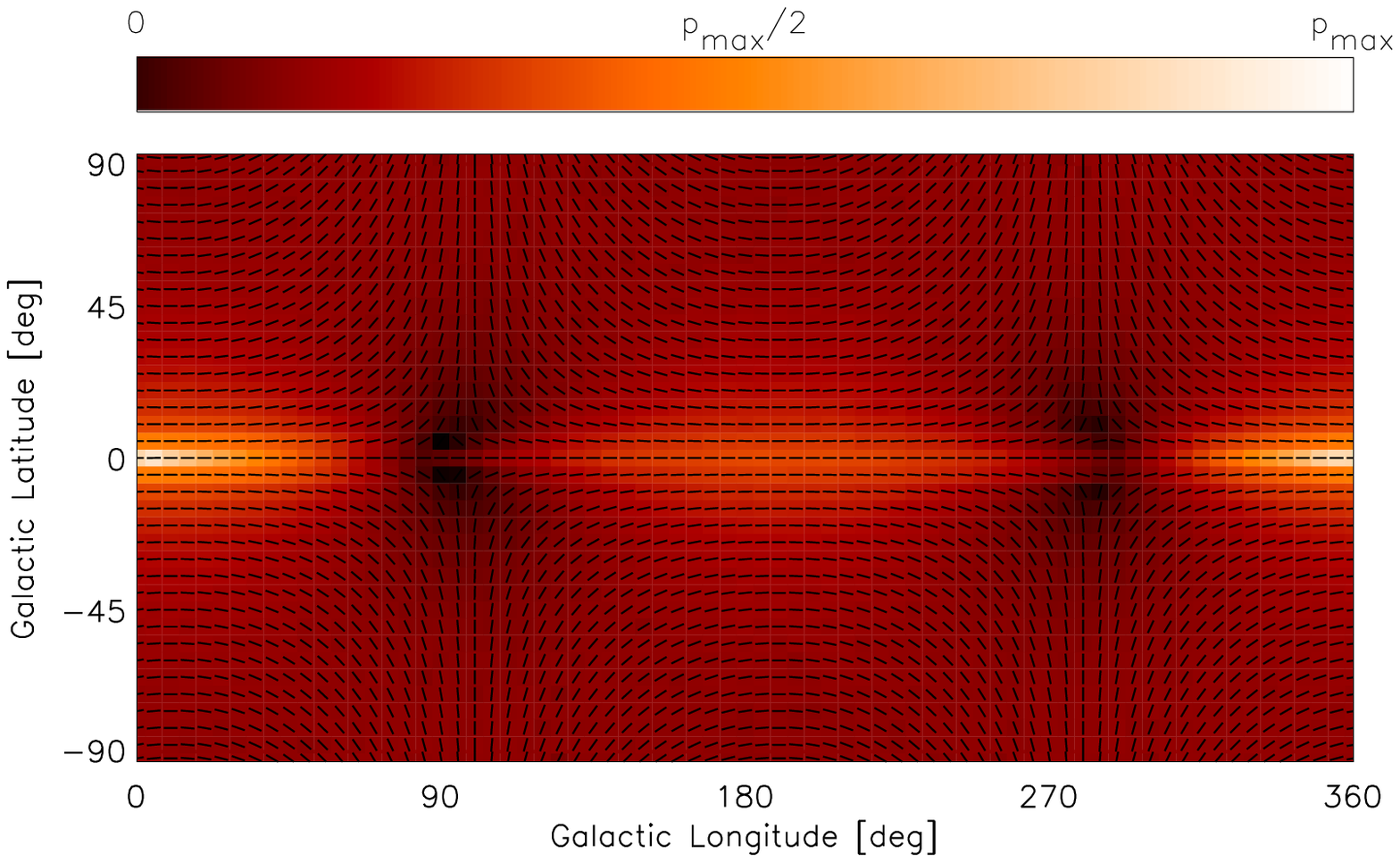}
	\centering
	\caption{\label{M513_DS}Same as Fig. \ref{predic_S0}, but for the DEHO Model 15.}
\end{figure}

\begin{figure}
	\includegraphics{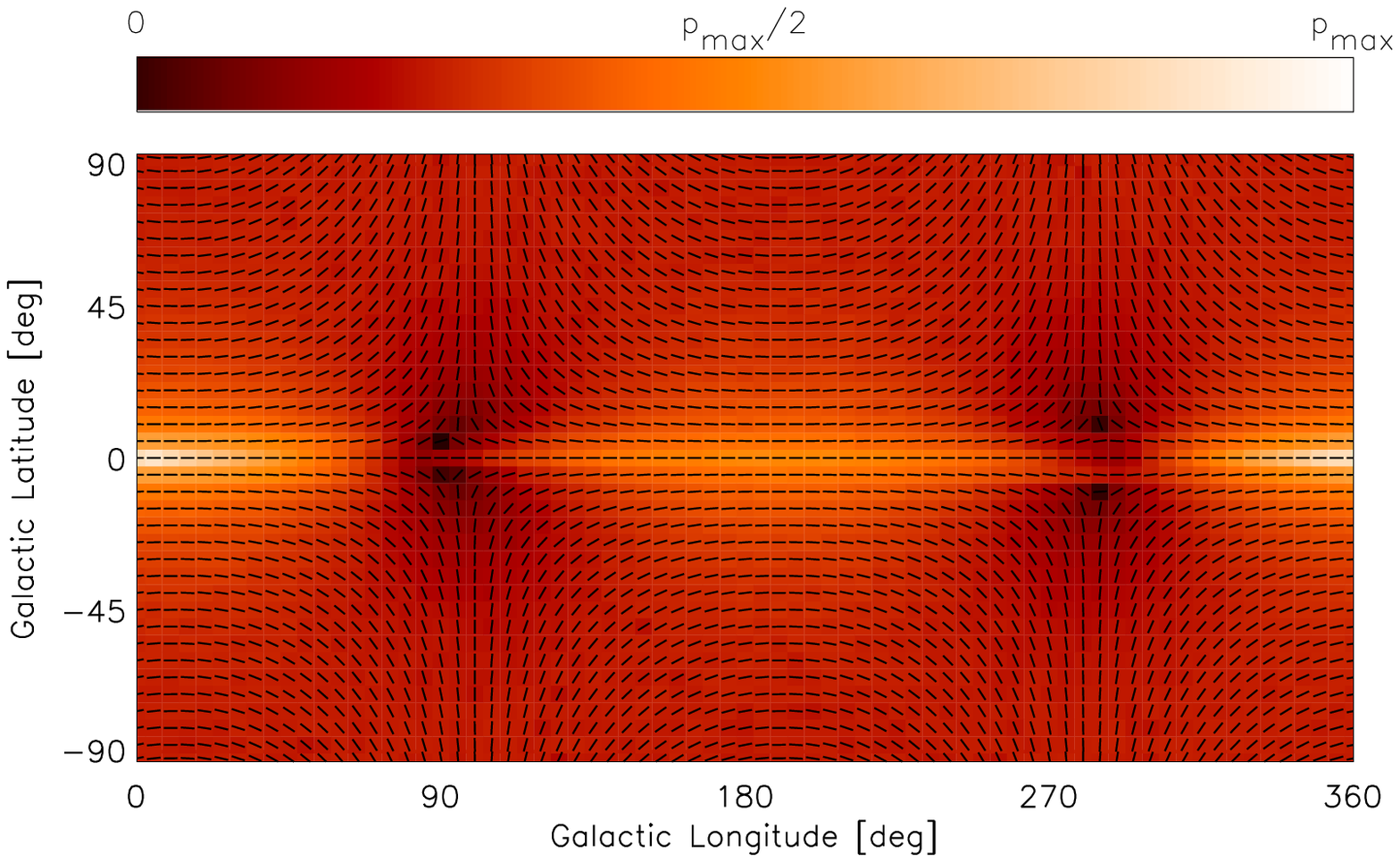}
	\centering
	\caption{\label{M513_SMB}Same as Fig. \ref{predic_S0}, but for the DEHO Model 16.}
\end{figure}

\begin{figure}
	\includegraphics{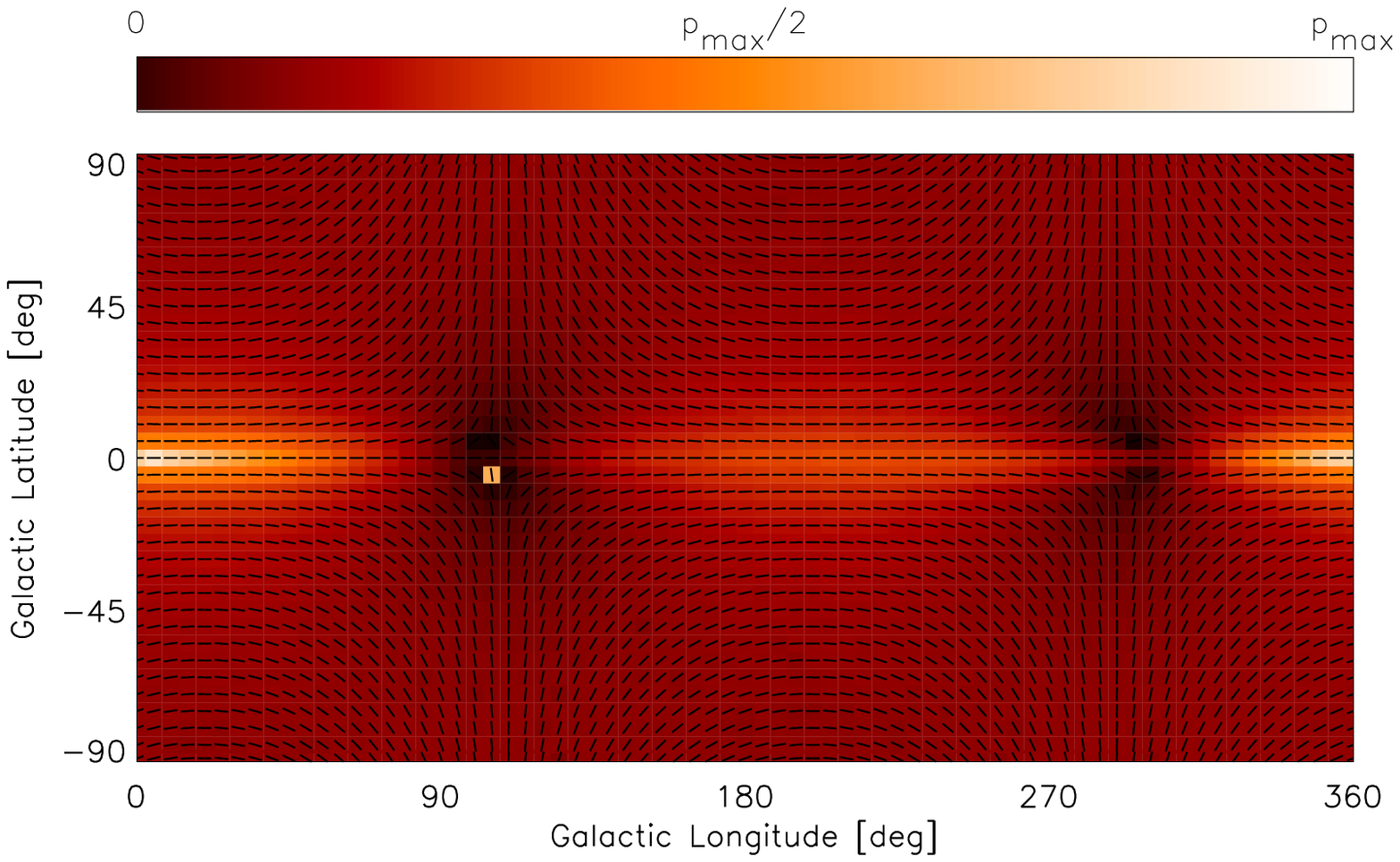}
	\centering
	\caption{\label{M527_DS}Same as Fig. \ref{predic_S0}, but for the DEHO Model 17.}
\end{figure}

\clearpage

\begin{figure}
	\includegraphics{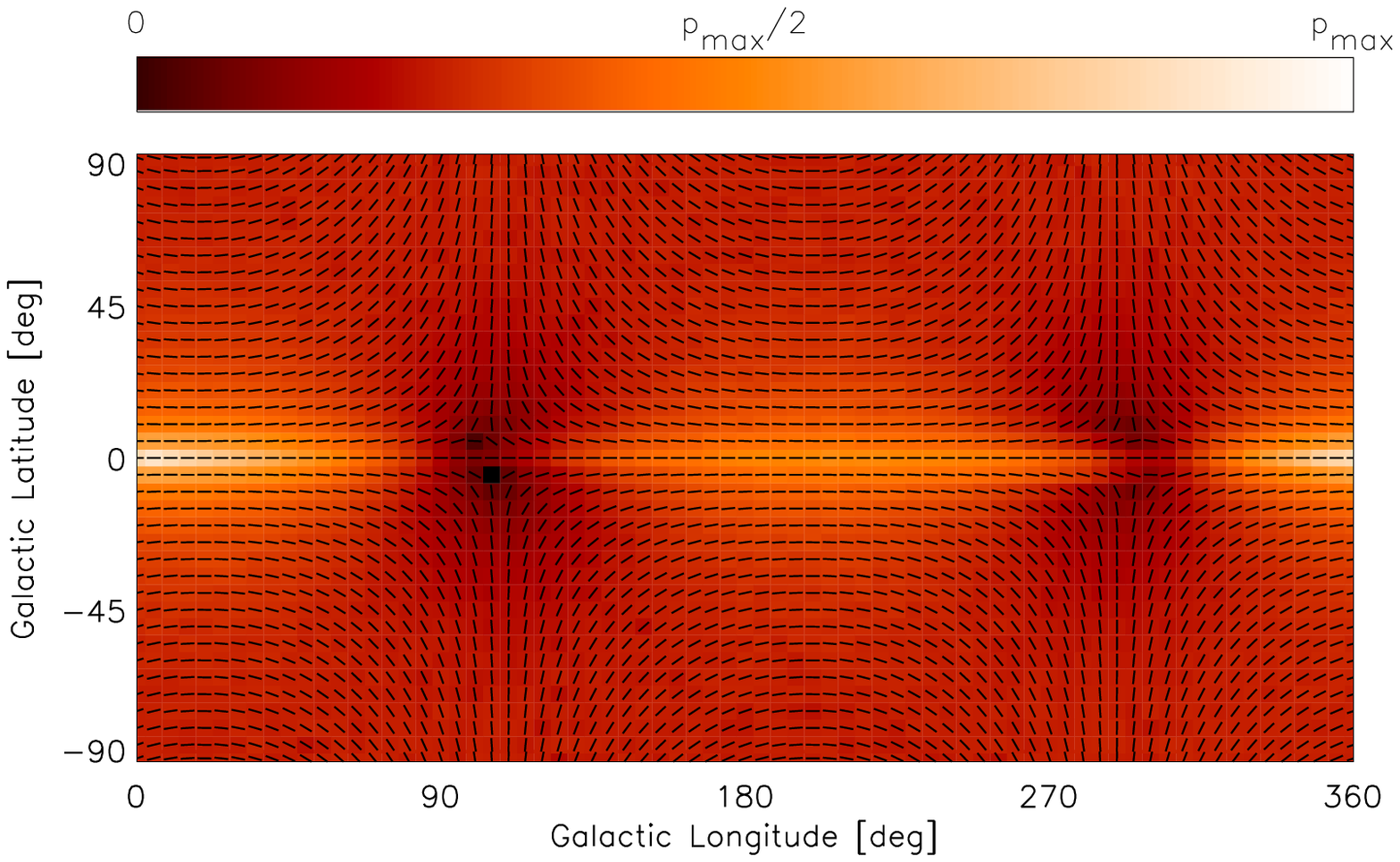}
	\centering
	\caption{\label{M527_SMB}Same as Fig. \ref{predic_S0}, but for the DEHO Model 18.}
\end{figure}

\begin{figure}
	\includegraphics{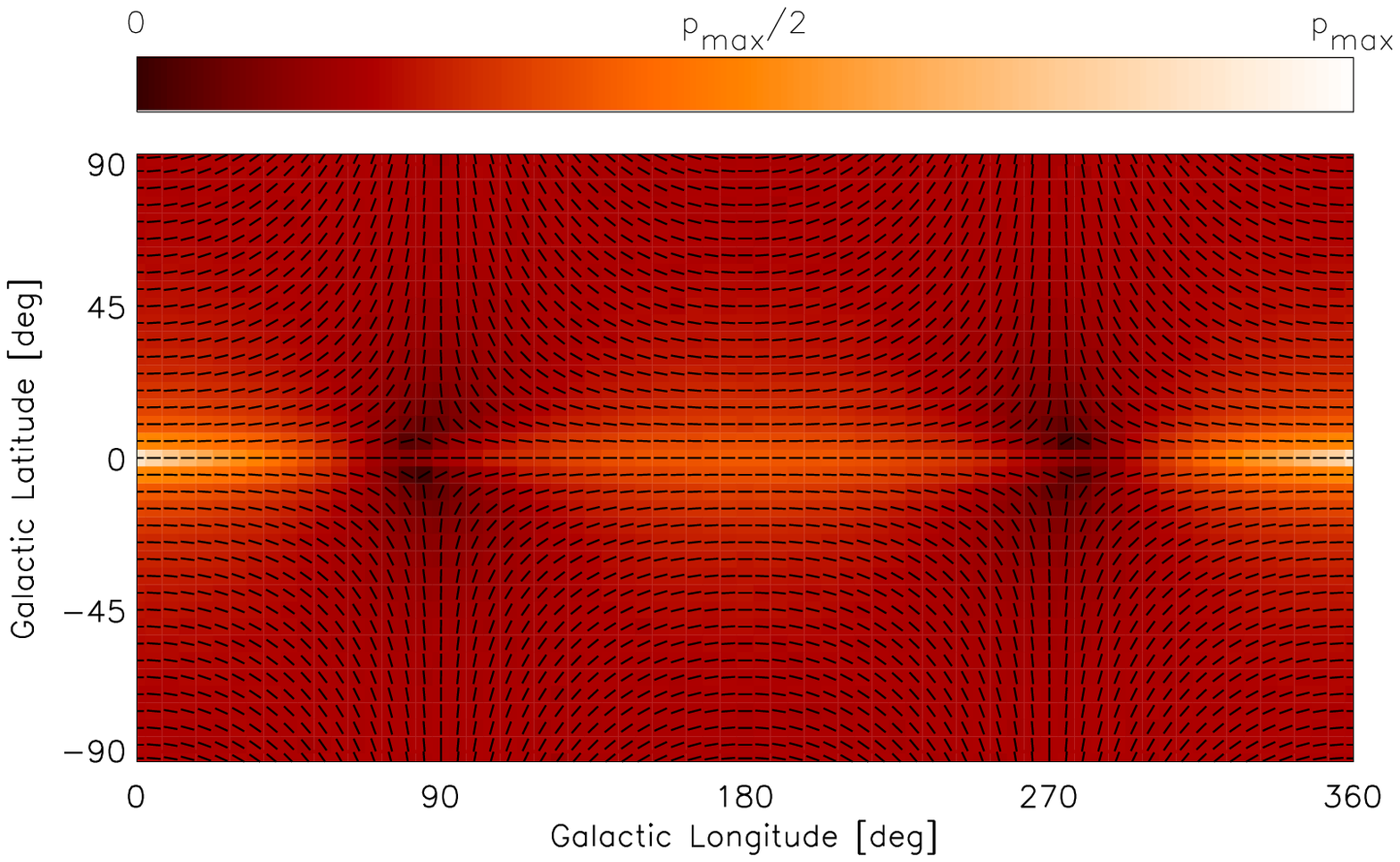}
	\centering
	\caption{\label{R0_DS}Same as Fig. \ref{predic_S0}, but for analytic axisymmetric Model 19.}
\end{figure}

\begin{figure}
	\includegraphics{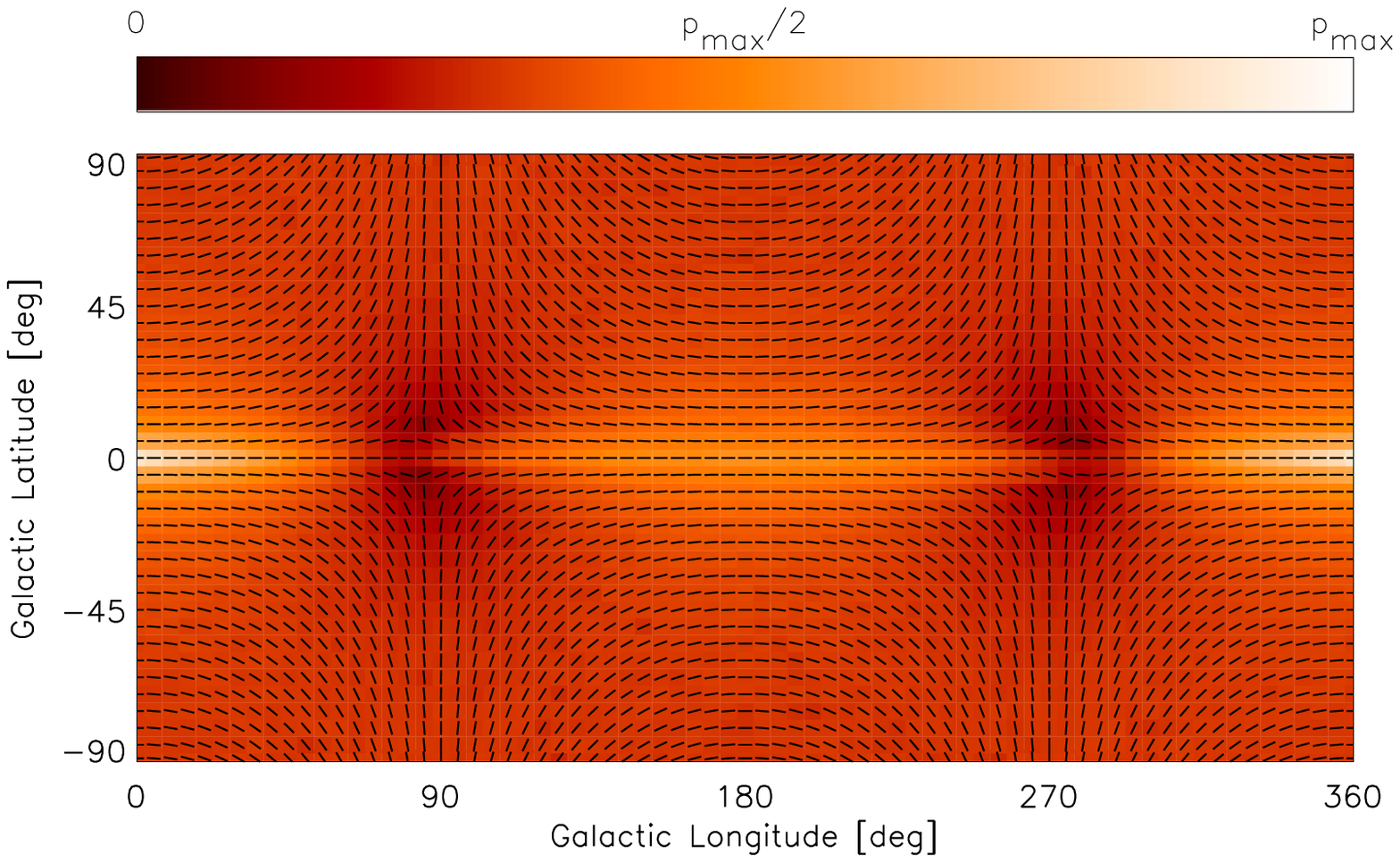}
	\centering
	\caption{\label{R0_SMB}Same as Fig. \ref{predic_S0}, but for analytic axisymmetric Model 20.}
\end{figure}

\begin{figure}
	\includegraphics{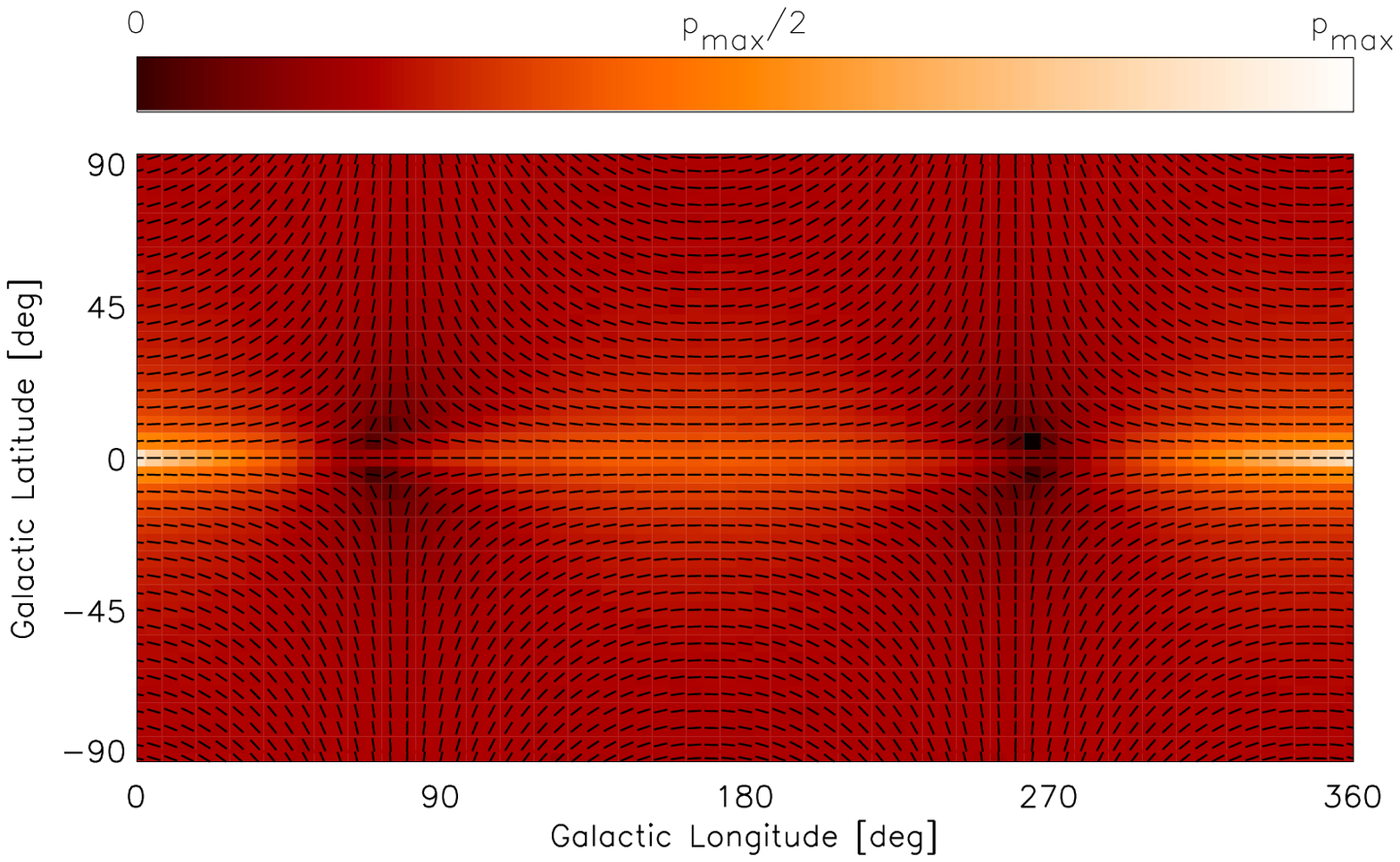}
	\centering
	\caption{\label{predic_AA}Same as Fig. \ref{predic_S0}, but for analytic axisymmetric Model 21.}
\end{figure}

\begin{figure}
	\includegraphics{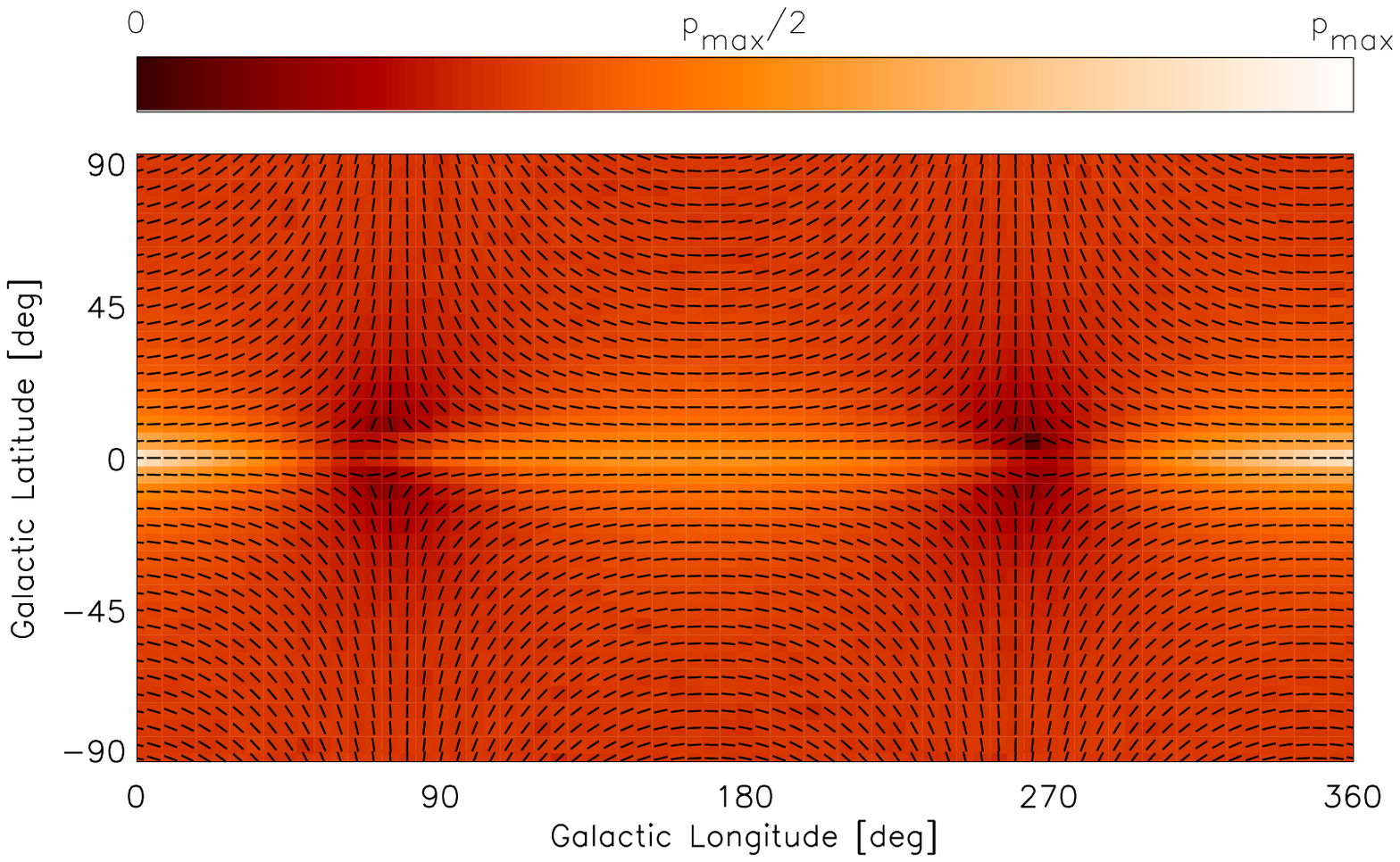}
	\centering
	\caption{\label{R115_SMB}Same as Fig. \ref{predic_S0}, but for analytic axisymmetric Model 22.}
\end{figure}

\begin{figure}
	\includegraphics{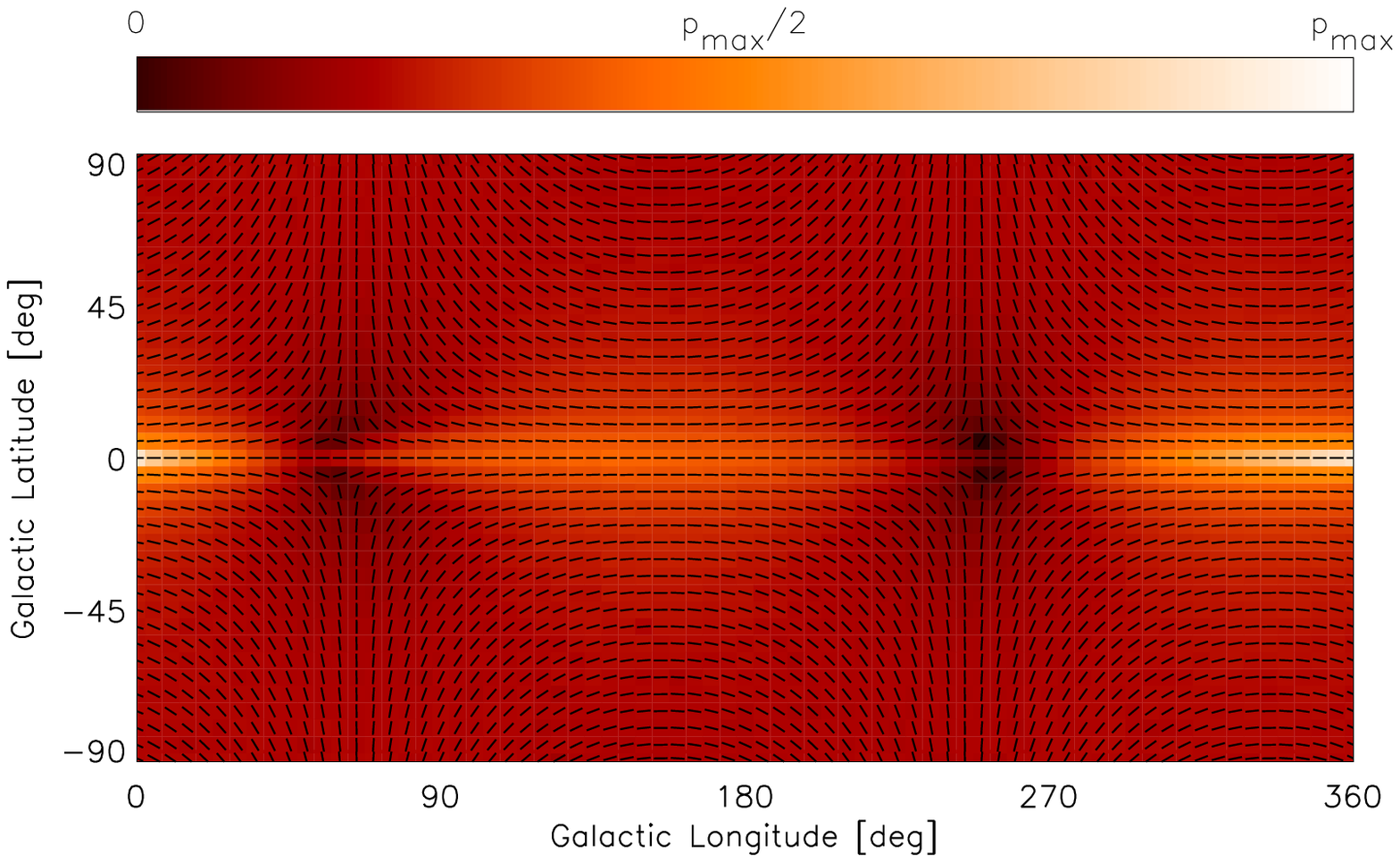}
	\centering
	\caption{\label{R24_DS}Same as Fig. \ref{predic_S0}, but for analytic axisymmetric Model 23.}
\end{figure}

\begin{figure}
	\includegraphics{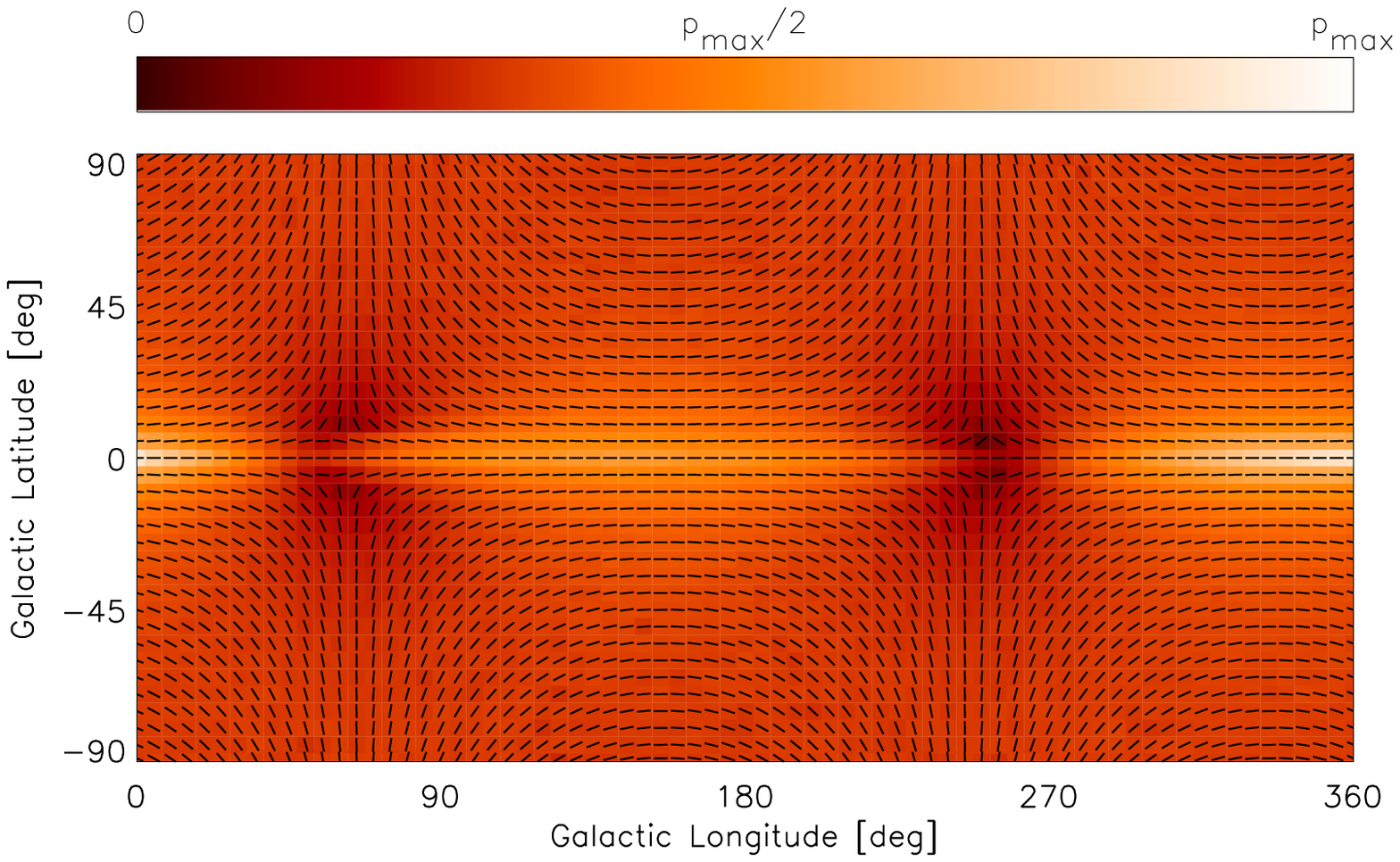}
	\centering
	\caption{\label{R24_SMB}Same as Fig. \ref{predic_S0}, but for analytic axisymmetric Model 24.}
\end{figure}
\end{landscape}

\begin{figure}
	\includegraphics[scale=1.5]{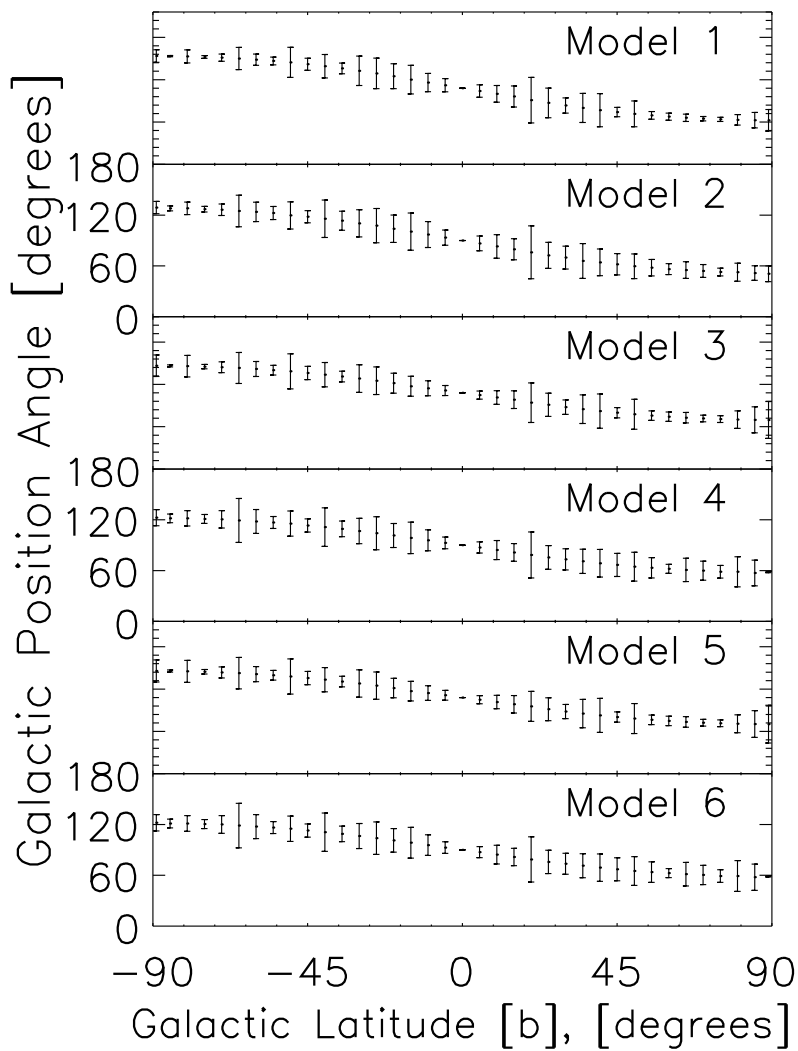}
	\centering
	\caption{\label{S0_cut}Predicted mean Galactic polarization PA in $10\arcmin \times 10\arcmin$ fields for simulated S0 models, from \citet{FS2000}, at $\ell=150\degr$, as a function of Galactic latitude. The error bars represent the $\pm 5\sigma$ PA dispersions for all stars in each Galactic latitude field. All of the S0 magnetic field models produce similar shapes.}
\end{figure}

\begin{figure}
	\includegraphics[scale=1.5]{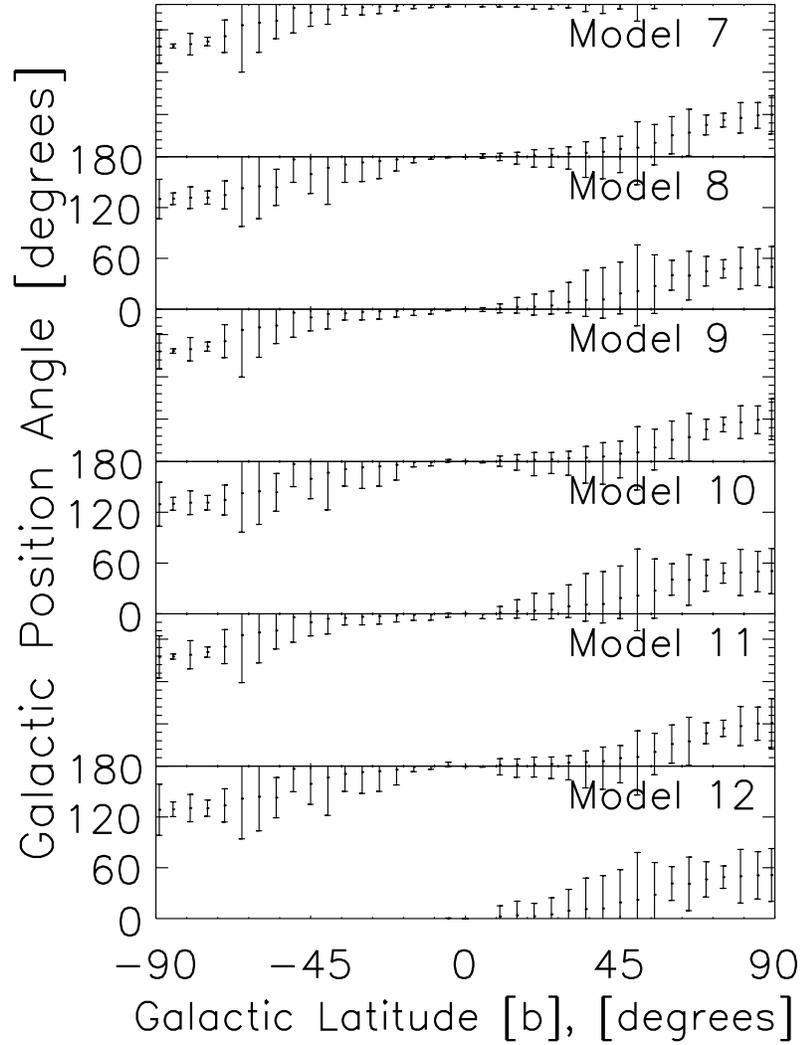}
	\centering
	\caption{\label{A0_cut}Similar to Fig. \ref{S0_cut}, but for A0 magnetic field models from \citet{FS2000}. Some data points appear twice because of the $180\degr$ ambiguity in PA.}
\end{figure}

\begin{figure}
	\includegraphics[scale=1.5]{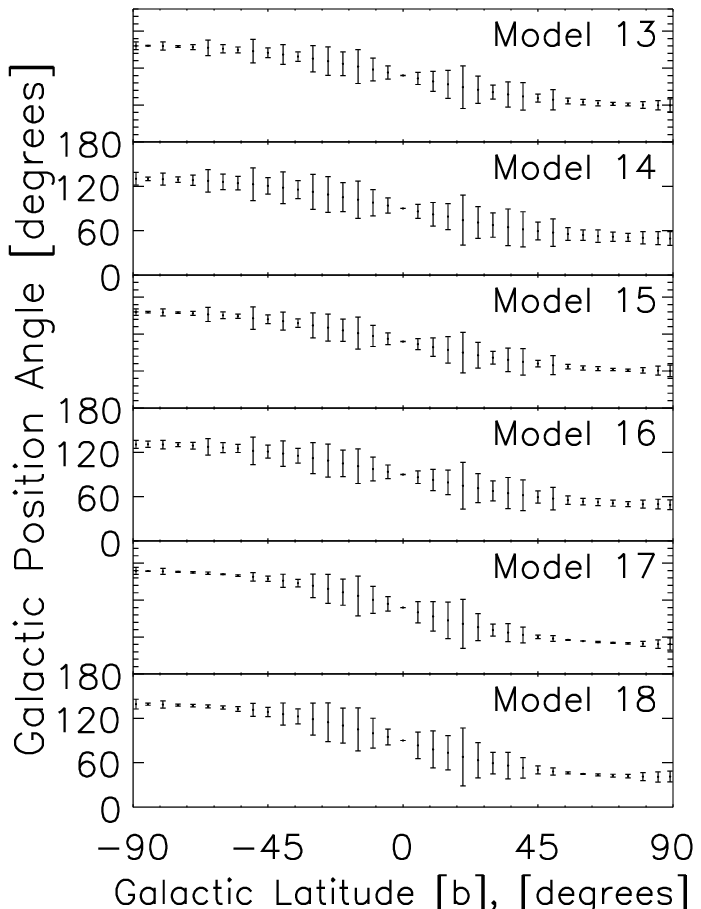}
	\centering
	\caption{\label{DEHO_cut}Same as Fig. \ref{S0_cut}, but for DEHO models from \citet{M10}.}
\end{figure}

\begin{figure}
	\includegraphics[scale=1.5]{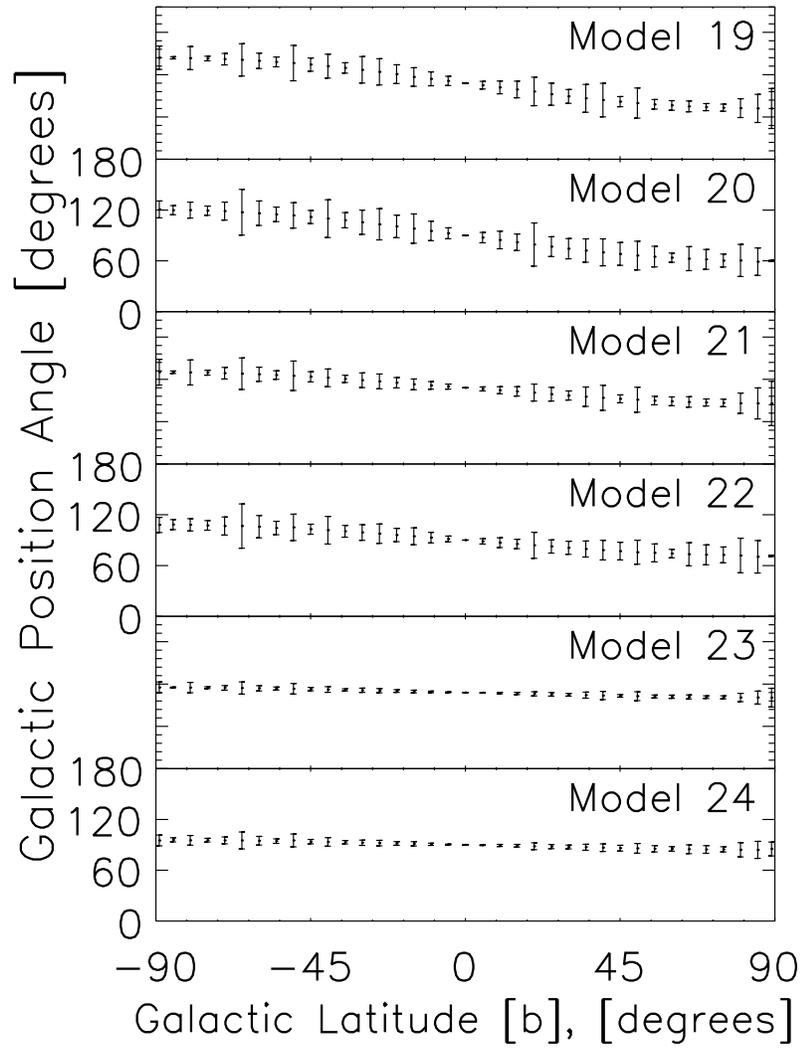}
	\centering
	\caption{\label{AA_cut}Same as Fig. \ref{S0_cut}, but for analytic axisymmetric models with different pitch angles.}
\end{figure}

\begin{figure}
	\includegraphics{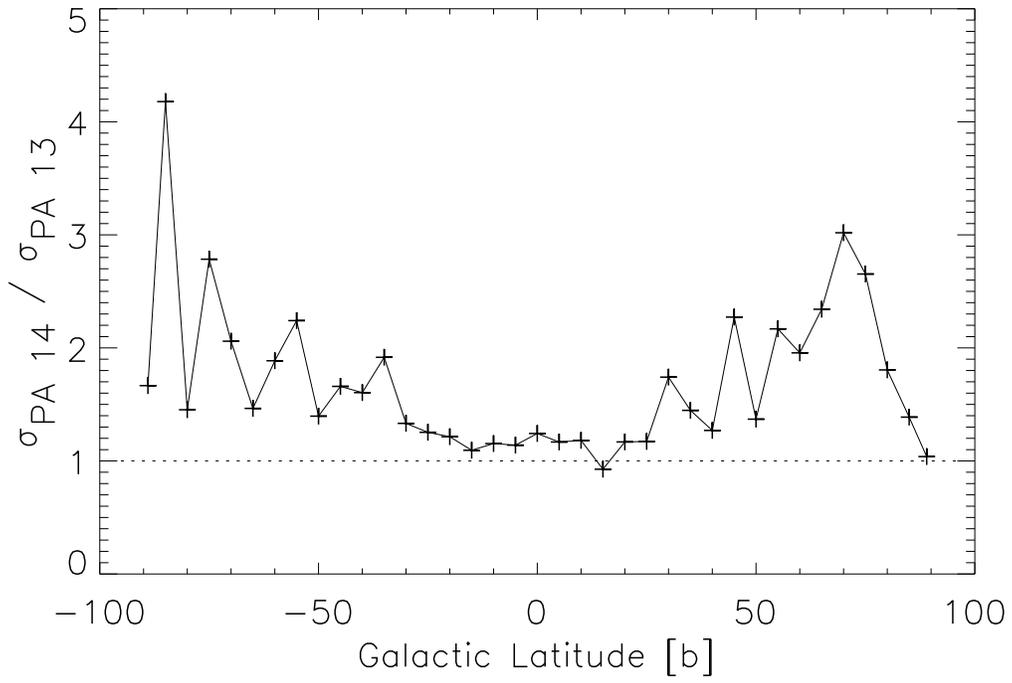}
	\caption{\label{dust_example} Ratio of Model 13 and Model 14 PA dispersions as a function of Galactic latitude for $\ell = 150\degr$.}
\end{figure}

\end{document}